\documentclass[ALICE,manyauthors]{etc/cernphprep}
\usepackage{etc/cernphprep}
\usepackage{etc/orcidlink}

\begin{document}

\begin{titlepage}
\PHyear{2022}   
\PHnumber{218}  
\PHdate{18 October} 

\title{Production of \kzero, \lmb (\almb), \Xis and \Oms in jets and in the underlying event in \pp and \pPb collisions}
\ShortTitle{Production of strange particles in jets and the UE in \pp and \pPb collisions}

\Collaboration{ALICE Collaboration\thanks{See Appendix~\ref{app:collab} for the list of collaboration members}}
\ShortAuthor{ALICE Collaboration} 

\begin{abstract}

The production of strange hadrons (${\rm K}^{0}_{\rm{S}}$, ${\Lambda}$, $\Xi^{\pm}$, and $\Omega^{\pm}$), baryon-to-meson ratios (${\Lambda}/{\rm K}^{0}_{\rm{S}}$, $\Xi/{\rm K}^{0}_{\rm{S}}$, and $\Omega/{\rm K}^{0}_{\rm{S}}$), and baryon-to-baryon ratios ($\Xi/{\Lambda}$, $\Omega/{\Lambda}$, and $\Omega/\Xi$) associated with jets and the underlying event were measured as a function of transverse momentum ($p_{\rm T}$) in pp collisions at $\sqrt{s} = 13$~TeV and p$-$Pb collisions at $\sqrt{s_{\rm NN}} = 5.02$~TeV with the ALICE detector at the LHC.
The inclusive production of the same particle species and the corresponding ratios are also reported.
The production of multi-strange hadrons, $\Xi^{\pm}$ and $\Omega^{\pm}$, and their associated particle ratios in jets and in the underlying event are measured for the first time.
In both pp and \pPb collisions, the baryon-to-meson and baryon-to-baryon yield ratios measured in jets differ from the inclusive particle production for low and intermediate hadron $p_{\rm T}$ (0.6$-$6~GeV/$c$).
Ratios measured in the underlying event are in turn similar to those measured for inclusive particle production.
In pp collisions, the particle production in jets is compared with $\textsc{Pythia} 8$ predictions with three colour-reconnection implementation modes.
None of them fully reproduces the data in the measured hadron $p_{\rm T}$ region.
The maximum deviation is observed for $\Xi^{\pm}$ and $\Omega^{\pm}$ which reaches a factor of about six.
The event multiplicity dependence is further investigated in p$-$Pb collisions.
In contrast to what is observed in the underlying event, there is no significant event-multiplicity dependence for particle production in jets.
The presented measurements provide novel constraints on hadronisation and its Monte Carlo description. In particular, they demonstrate that the fragmentation of jets alone is insufficient to describe the strange and multi-strange particle production in hadronic collisions at LHC energies.

\end{abstract}
\end{titlepage}

\setcounter{page}{2} 



\section{Introduction}%
\label{sec:Introduction}

High-energy heavy-ion collisions at the LHC create a hot and dense form of matter called quark$-$gluon plasma (QGP)~\cite{Shuryak:1983ni,Cleymans:1985wb,Bass:1998vz,Satz:2000bn,Jacak:2012dx,Muller:2012zq}.
The droplet of QGP created in the collision rapidly expands as a strongly-coupled liquid and cools down until a temperature near the phase transition at which the deconfined partons hadronise into ordinary colour-neutral matter~\cite{Borsanyi:2010cj,Bhattacharya:2014ara,Braun-Munzinger:2015hba}.
Systematic studies of particle production, transverse momentum spectra and correlations of identified particles allow to investigate the properties of the partonic phase and the hadronisation process itself.

The interpretation of heavy-ion results (with hot nuclear matter effects) and extraction of the QGP properties require studies of particle production in small collision systems, proton--proton (pp) and proton--nucleus (pA).
Previously, the measurements in these small collision systems were thought of as a necessary foundation to quantify the initial and final state effects of the so-called cold nuclear matter.
However, during the last decade, the study of small systems has gained increased interest as a research field in its own right.
In particular, similar effects as those present in heavy-ion collisions have been observed in \pp and \pPb collisions where the formation of a QGP was not expected ~\cite{Aad:2015gqa, Abelev:2012ola, ABELEV:2013wsa, Khachatryan:2015waa, Acharya:2019vdf, Abelev:2014uua, ALICE:2015mpp}.
These include, for example, the long-range angular correlations~\cite{Aad:2015gqa, Abelev:2012ola, ABELEV:2013wsa} and the non-vanishing elliptic flow coefficient measured using multi-particle cumulant analyses~\cite{Acharya:2019vdf, Khachatryan:2015waa}.
The magnitude of these effects increases smoothly with system size and particle multiplicity from pp, \pPb to \PbPb collisions.
Another new feature observed in high-multiplicity \pp and \pPb collisions is the enhancement of the baryon-to-meson yield ratios, p$/\pi$ and $\Lambda/\kzero$, at intermediate transverse momentum $\pT$ ($2$--$6$~\GeVc)~\cite{Acharya:2018orn, Khachatryan:2016yru, ALICE:2013cdo, ALICE:2017jyt}, which is qualitatively similar to that observed in \PbPb collisions.
Moreover, the strange to non-strange hadron ratio increases continuously as a function of charged-particle multiplicity density from low-multiplicity \pp to high-multiplicity \pPb collisions to eventually reach the values observed in \PbPb collisions~\cite{ALICE:2013wgn, ALICE:2017jyt, Khachatryan:2016yru}.
These findings suggest the possible existence of a common underlying mechanism which would determine the chemical composition of particles produced from small to large collision systems.
On the other hand, measurements of jet production at midrapidity in small systems do not exhibit nuclear modifications~\cite{Acharya:2019jyg, Acharya:2019tku, ALICE:2014dla, Abelev:2013fn, Acharya:2018eat, Acharya:2017okq, Adam:2015xea, ALICE:2016faw}.

The enhancement of baryon-to-meson yield ratios in the intermediate $\pT$ region has been related to the interplay of radial flow and parton recombination~\cite{Fries:2003vb, Bozek:2011gq, Muller:2012zq}.
In a recent study, the ALICE Collaboration investigated baryon-to-meson yield ratios in two separate parts of the event -- inside a jet and in the event portion perpendicular to the jet cone -- in \pp collisions at $\s = 7$ and $13$~\TeV and \pPb collisions at \fivenn~\cite{Acharya:2021oaa,ALICE:2021nvv}.
The results show that the enhancement of the $\lmb/\kzero$ ratio at intermediate $\pT$ obtained from inclusive particle production measurements in \PbPb collisions at \twosevensix~\cite{ALICE:2013cdo} and in high-multiplicity  \pp collisions~\cite{ALICE:2020jsh} is not present in the low-$z$ fragmentation products of jets.
In addition to these effects, one can expect that particle production in this $\pT$ region results from the hard fragmentation of partons of $\pT$ in the $4$--$8$~\GeVc range (momentum fraction $z = \pT^{\rm hadron} / \pT^{\rm parton} \approx 0.5$).
This is due to the steeply falling power-law spectrum characteristic for parton production. This so-called ``leading particle effect'' was described in terms of a ``trigger bias''~\cite{Ellis:1976hun}.
Studying the yield ratios of particles associated with jets allows us to explore a larger $z$ range, providing new constraints on whether the baryon-to-meson yield ratio enhancement originates from jet fragmentation.

In this article, the $\pT$-differential baryon-to-meson and multi-strange-to-strange particle ratios are studied in jets reconstructed using the charged-particle component (charged-particle jets) and in the underlying event associated to jets.
This provides further understanding of the contribution of soft and hard processes to the enhancement of the baryon-to-meson yield ratios at intermediate $\pT$ and of the strange particle yields as a function of multiplicity in small systems.
In particular, the measurement of the production of \kzero, \lmb (\almb), \Xis, and \Oms in charged-particle jets and in the underlying event in \pp collisions at \thirteen and \pPb collisions at \fivenn is reported.
Strange particles are reconstructed in the pseudorapidity range $|\eta| < 0.75$.
Jets are reconstructed with a transverse momentum $\pTjch > 10$~\GeVc and in the pseudorapidity range $\abs{\etaJ} < 0.35$ with a resolution parameter $R = 0.4$ (referred to as jet radius in the following).
The strange particles produced inside a jet are characterised as a function of the distance between the particle momentum vector and the jet axis in the $\eta$--$\varphi$ plane, where $\varphi$ is the azimuthal angle.

The results presented in this article surpass the precision of the previous ALICE $\pT$-differential measurements~\cite{Acharya:2021oaa} both in \pp at $\sqrt{s}=7$~\TeV and \pPb at \fivenn collisions.
The studies are extended to the multi-strange sector and the charged-particle multiplicity dependence is investigated as well.
The baryon-to-meson and baryon-to-baryon yield ratios inside jets are compared to the same ratios obtained from inclusive events and the underlying event.
Results measured in pp collisions are compared with \Pyeight~\cite{Sjostrand:2014zea} simulations.

The article is structured as follows.
In Section~\ref{sec:Detector}, the ALICE apparatus and the data samples used for the analysis are presented.
In Section~\ref{sec:Analysis}, the methods adopted for charged-particle jet reconstruction, strange particle reconstruction, and particle--jet matching are described.
This section also includes the estimate of the associated systematic uncertainties.
The measurement of strange hadron $\pT$ distributions and the corresponding yield ratios, together with their comparison with model predictions, are presented and discussed in Section~\ref{sec:Results}.
The paper is summarized in Section~\ref{sec:Summary}.

\section{ALICE detector and data selection}%
\label{sec:Detector}

The ALICE apparatus and its performance are described in Refs.~\cite{Collaboration_2008, Abelev:2014ffa}.
This analysis mainly relies on the central barrel tracking system and the forward \VZERO detector~\cite{Abbas:2013taa}.
The central barrel detectors used for this analysis are the Inner Tracking System (ITS)~\cite{Aamodt:2010aa}, the Time Projection Chamber (TPC)~\cite{Alme:2010ke}, and the Time-Of-Flight detector (TOF)~\cite{Akindinov:2004cj, Akindinov:2009zze, Akindinov:2010zzb}.
These detectors cover the pseudorapidity region $|\eta| < 0.9$ and are located inside a large solenoidal magnet providing a 0.5~T magnetic field.

The ITS, the innermost barrel detector, consists of six cylindrical layers of high spatial resolution silicon detectors using three different technologies.
The two innermost layers~(Silicon Pixel Detector, SPD) are based on silicon pixel technology and cover $|\eta| < 2.0$ and $|\eta| < 1.4$, respectively.
The SPD is used to reconstruct the primary vertex of the collision and short track segments, which are called "tracklets".
The four outer ITS layers consist of silicon drift (SDD) and strip (SSD) detectors, with the innermost (outermost) layer having a radius $r = 15$ $(43)$~cm.
The SDD and SSD are able to measure the specific ionization energy loss (\dEdx) with a relative resolution of about $10\%$ in the low-$\pT$ region (up to $\sim 1$~\GeVc)~\cite{Aamodt:2010aa}.
The ITS is also used to reconstruct and identify low-momentum particles down to $100$~\MeVc that cannot reach the TPC.

The TPC is a large cylindrical gaseous detector filled with a Ne-CO$_{2}$ gas mixture.
The radial and longitudinal dimensions of the TPC are about $85 < r < 250 $~cm and $-250 < z < 250 $~cm, respectively.
As the main tracking device, the TPC provides full azimuthal acceptance for tracks in the region $|\eta| < 0.9$.
In addition, it provides charged-hadron identification via the \dEdx measurement.
At low $\pT$, the \dEdx resolution of $5.2\%$ for a minimum ionizing particle allows track-by-track particle identification~\cite{Alme:2010ke}.
On the other hand, at intermediate and high $\pT$ ($\gtrsim 2.0$~\GeVc), the energy loss distributions of different particle species start to overlap. Therefore, from there on, particles have to be statistically separated via a multi-Gaussian fit to the \dEdx distributions.

The TOF, located at a radius of 3.7 m, outside of the TPC, measures the flight time of the particles.
It consists of a cylindrical array of multi-gap resistive plate chambers with an intrinsic time resolution of $50$~ps.
It covers the range $\abs{\eta} < 0.9$ with full azimuthal acceptance.
It can provide particle identification over a broad $\pT$ range ($0.5 \lesssim \pT \lesssim 2.7$~\GeVc).
The total time-of-flight resolution, including the collision time resolution, is about $90$~ps in \pp and \pPb collisions~\cite{ALICE:2020jsh}.
The \VZERO detector, composed of two scintillator arrays, V0A (covering a pseudorapidity range of $2.8 < \eta < 5.1$) and V0C ($-3.7 < \eta < -1.7$), is utilized for triggering and event classification based on charged-particle multiplicity.

Data from \pp collisions at \thirteen and from \pPb collisions at \fivenn are used in this analysis.
The \pp and \pPb data samples were recorded with the ALICE detector in $2016$--$2017$ and $2016$, respectively.
These data were collected with a minimum bias (MB) trigger requiring at least one hit in both \VZEROA and \VZEROC in coincidence with the bunch crossing.

Interaction vertices are reconstructed by the extrapolation of ITS tracklets towards the average beam line.
Pileup events, due to multiple interactions in the triggered bunch crossing, are removed by exploiting the correlation between the number of SPD hits and tracklets.
The coordinate of the primary vertex along the beam direction is required to be within $\pm 10$~cm with respect to the nominal position of the ALICE interaction point.
After event selection, the \pp sample consists of $1.5$ billion events.
The integrated luminosity of $\lumi_{\rm int} = 9.38\pm 0.47$~nb$^{-1}$ based on the visible cross section observed by the V0 trigger was extracted from a van der Meer scan~\cite{ALICE:2021leo}.
About $500$ million events from the \pPb samples were selected, which correspond to an integrated luminosity of $\lumi_{\rm int} = 295 \pm 11 $~$\mu{\rm b}^{-1}$~\cite{ALICE:2019oyn}.
The \pPb events are divided into three multiplicity classes based on the total charge deposited in the V0A (in the Pb-going direction).
The multiplicity intervals and their corresponding mean charged-particle density (\dndeta) measured at midrapidity ($\abs{\eta} < 0.5$) are given in Ref.~\cite{Adam:2015pza}.

\section{Analysis}%
\label{sec:Analysis}

\subsection{Charged-particle jet reconstruction}%
\label{sec:JetRec}

The charged-particle jets are reconstructed using the FastJet package~\cite{Cacciari:2011ma} with the \akT algorithm~\cite{Cacciari:2008gp} with a resolution parameter $R = 0.4$.
As the inputs, the charged particles are reconstructed using the ITS and TPC information.
Tracks with $\pT > 0.15$~\GeVc are accepted over the pseudorapidity range $\abs{\eta_{\rm trk}} < 0.9$ and with azimuthal angle $0 < \varphi < 2\pi$.

The reconstructed jet axis pseudorapidity is required to be in the range $\abs{\etaJ} < 0.35$.
This condition ensures that the jet cone is fully contained within the $\eta$-acceptance for strange particles.
A selection on the charged-particle jet $\pT$, $\pTj^{\rm ch} > 10$~\GeVc, is applied to ensure that the jet originates from the hard scattering process~\cite{Acharya:2021oaa}.

In a hadron--hadron collider event, the two outgoing partons from the hard scattering are accompanied by particles that arise, e.g. from multiple parton interactions, which form a background for the jet production measurement.
In pp collisions, the $\pT$ density per unit area in the $\eta$--$\varphi$ plane~($\rhobkg^{\rm ch}$) of this background is determined from the $\kT$ algorithm~\cite{Catani:1993hr, Ellis:1993tq} to be around $1$~\GeVc~rad$^{-1}$, which is negligible and, hence, not subtracted in this analysis.
The background density in events with at least one jet in $\pT > 10 $~\GeVc is around $3$~\GeVc~rad$^{-1}$ in \pPb collisions, which is more significant than pp collisions.
Hence the reconstructed \pT\ of the jet is corrected for the background contribution~\cite{Cacciari:2007fd} using the formula
\begin{equation}
\pTj^{\rm ch} = \pTj^{\rm rec} - \rhobkg^{\rm ch}\times\Ajet,
\label{eq:DeltaPt}
\end{equation}
where $\pTj^{\rm rec}$ is the reconstructed jet $\pT$ and $\Ajet$ is the jet area.
$\Ajet$ is calculated by the active ghost area method of FastJet, with a ghost area of $0.005$~\cite{Cacciari:2008gn}.
The estimation of the background density $\rho_{\rm bkg}^{\rm ch}$ in sparse systems such as \pPb collisions is based on the method described in Ref.~\cite{Chatrchyan:2012tt}. This method allows to circumvent problems arising from the use of ghost jets applicable to larger collision systems~\cite{Chatrchyan:2012tt}. Under this method, empty areas are instead accounted for by applying a correction factor to the background density as follows
\begin{equation}
\rhobkg^{\rm ch} = C\times{\rm median}\left\{\frac{p_{\rm T,jet}^{\rm rec}}{A_{\rm jet}}\right\}, ~ {\rm with}~ C = \frac{\sum_{i}A_{i}}{A_{\rm acc}},
\label{eq:RhoCMS}
\end{equation}
where $A_{i}$ is the area of each $\kT$ jet with at least one real track, i.e. excluding ghosts and $A_{\rm acc}$ is the area of the charged-particle acceptance, namely $(2 \times 0.9) \times 2\pi$.
The background estimate is made more accurate by excluding the two clusters with the largest $\pT$ from the $\rho_{\rm bkg}$ calculation as given by Eq.~(\ref{eq:RhoCMS}).

\begin{figure}[!t]
\begin{center}
\includegraphics[width=.49\textwidth]{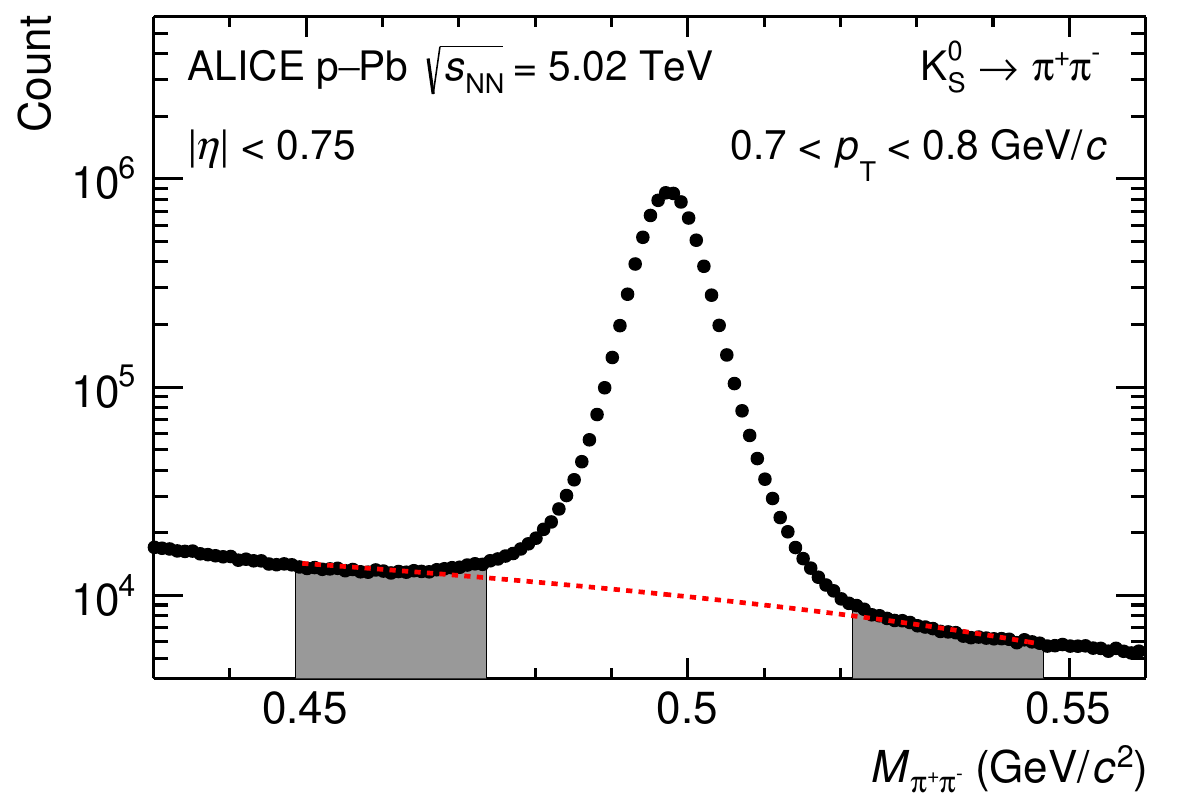}
\includegraphics[width=.49\textwidth]{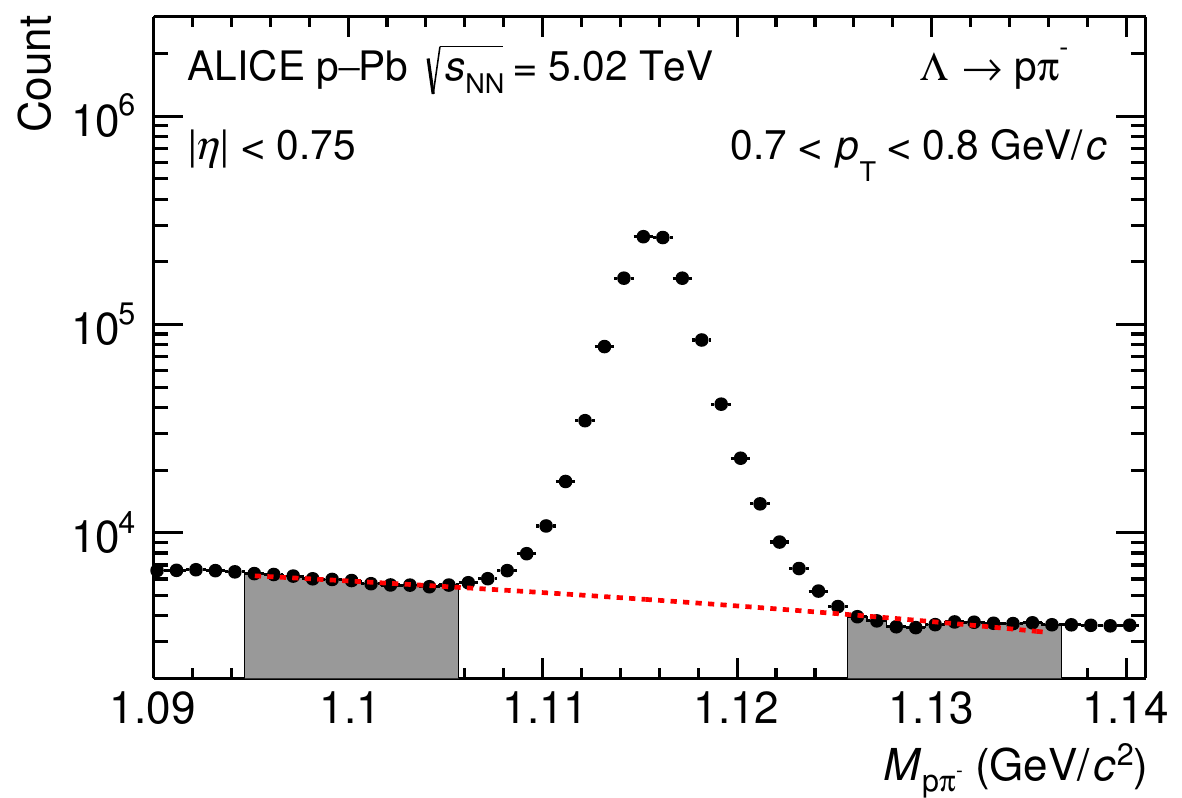}
\includegraphics[width=.49\textwidth]{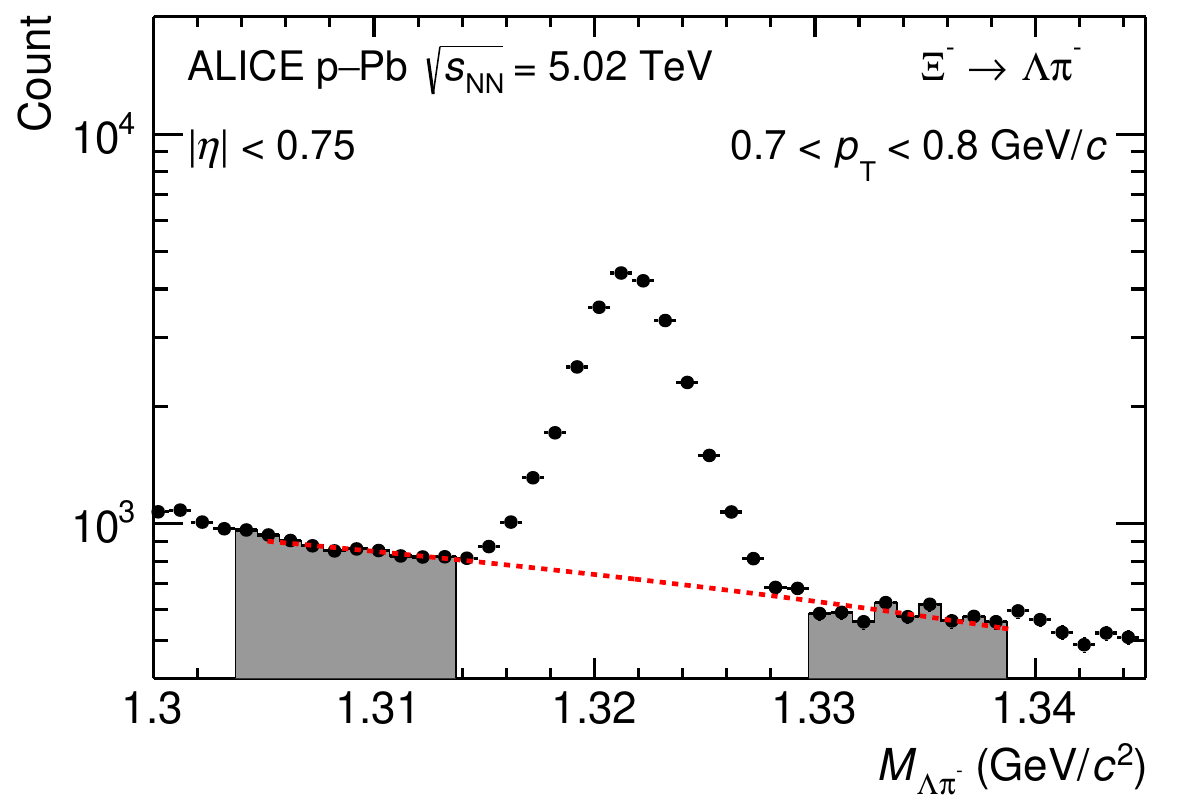}
\includegraphics[width=.49\textwidth]{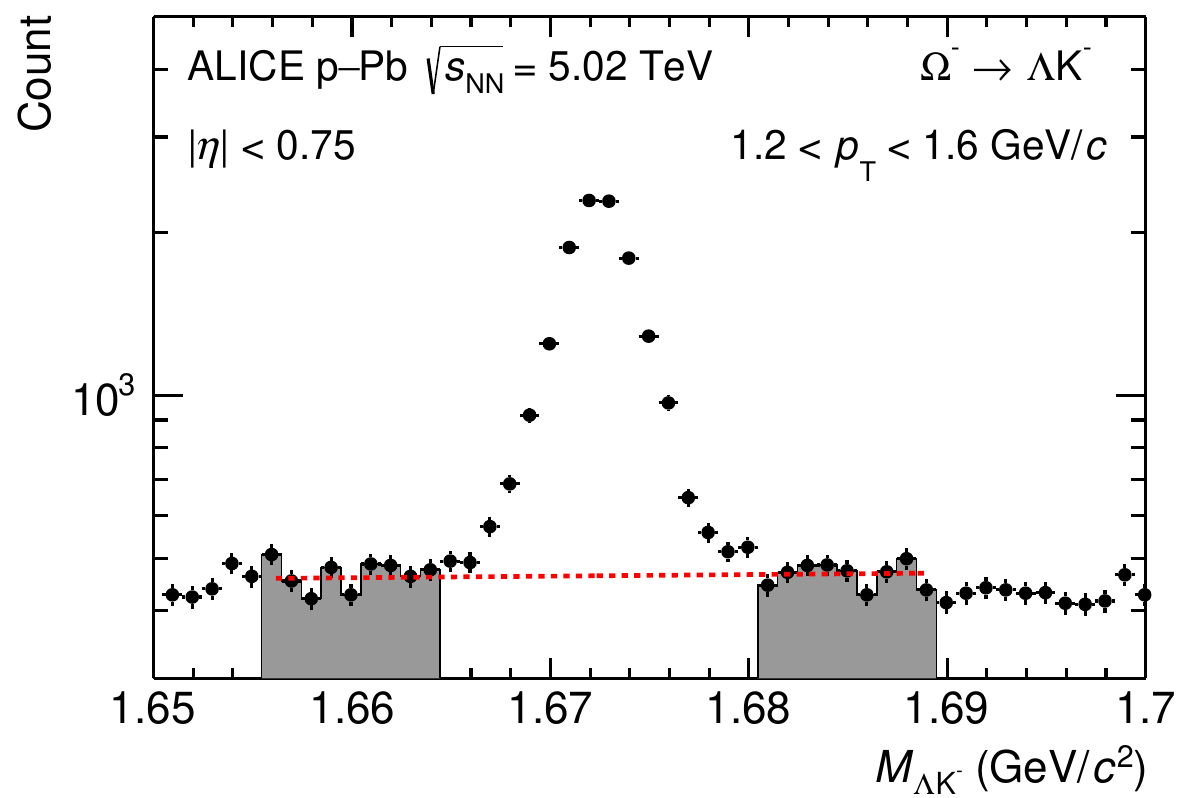}
\caption{Invariant mass distribution for $\kzero$, $\lmb$, $\X$, and $\Om$ in different $\pT$ intervals in MB \pPb collisions at \fivenn.
The candidates are reconstructed in $|\eta|<0.75$. The grey areas are used to determine the background (red dashed lines), see text for details.}
\label{fig:InvM}
\end{center}
\end{figure}

\subsection{Strange particle reconstruction}%
\label{sec:ParRec}

The strange particles \kzero, \lmb, \almb, \Xis, and \Oms are measured at midrapidity ($\abs{\eta} < 0.75$) via the reconstruction of their specific weak decay topology.
The following charged decay channels with the corresponding branching ratios~($B. R.$)~\cite{ParticleDataGroup:2022pth} are used:
$$
\begin{aligned}
\kzero      & \to \pip + \pim               & B.R. & = (69.20 \pm 0.05) \%, \\
\lmb (\almb) & \to \mathrm{p}(\pbar) + \pim (\pip)    & B.R. & = (63.9  \pm 0.5)  \%, \\
\X (\Ix)    & \to \lmb (\almb) + \pim (\pip) & B.R. & = (99.887 \pm 0.035) \%, \\
\Om (\Mo)   & \to \lmb (\almb) + \kam (\kap) & B.R. & = (67.8  \pm 0.7)  \%. \\
\end{aligned}
$$
The proton, pion, and kaon tracks (daughter tracks) are identified via their measured energy deposition in the TPC~\cite{Abelev:2014ffa}.
The identification of the \Vzero candidates (\kzero and \lmb (\almb) that decay into two oppositely charged daughter particles) and cascade candidates (\Xis and \Oms that decay into a ``bachelor'' charged meson, identified as $\pi^{\pm}$ or ${\rm K}^{\pm}$, plus a \Vzero decaying particle, giving the cascade decay topology) follow those presented in earlier ALICE publications~\cite{Aamodt:2011zza,Abelev:2012jp,Acharya:2018orn,ALICE:2013wgn,ALICE:2020jsh,ALICE:2019avo}.
In addition, the contributions of pileup collisions outside the trigger bunch crossing (``out-of-bunch pileup'') are removed.
This is achieved by requiring that at least one of the tracks corresponding to charged particle decays matches a hit in a ``fast'' detector (either the ITS or the TOF detector).
The selections in this analysis are summarized in Tables~\ref{tab:V0Cut},~\ref{tab:CascadeCut} in Appendix~\ref{app:cuts}.

The signal extraction is performed as a function of $\pT$.
The invariant mass distribution in each $\pT$ interval is fitted with a Gaussian function for the signal and a linear function for the combinatorial background.
Examples of the invariant mass distribution fits for all particles are shown in Fig.~\ref{fig:InvM}.
This allows for the extraction of the mean ($\mu$) and width ($\sigma$) of the signal.
The ``peak'' region is defined as that within $\pm 6\sigma$ for \Vzeros and $\pm 3\sigma$ ($\pm 4\sigma$) for cascades in \pp (and \pPb) collisions with respect to $\mu$ for each $\pT$ interval.
The ``background'' regions are defined on both sides of the peak region (see the gray areas in Fig.~\ref{fig:InvM}).
The $\pT$-differential yields of strange particles are obtained by subtracting the integral of the background fit function in the peak region from the total bin count in the same region (see Ref.~\cite{Aamodt:2011zza} for the details).

\subsection{Matching of strange particles to jets}%
\label{sec:ParJetMatch}

The strategy for obtaining strange hadrons associated to hard scatterings, selected by charged-particle jets (JE particles), follows that presented in Ref.~\cite{Acharya:2021oaa}.
Particles are defined as located inside the jet cones (JC) if their distance to the jet axis in the $\eta$--$\varphi$ plane
\begin{equation}\label{eq:defRpj}
R({\rm particle, \, jet}) = \sqrt{(\eta_{\rm particle} -\eta_{\rm jet})^{2} + (\varphi_{\rm particle} -\varphi_{\rm jet})^{2}}
\end{equation}
is less than a given value $R_{\max}$,
\begin{equation}\label{eq:defJC}
R({\rm particle, \, jet}) < R_{\max},
\end{equation}
where $R_{\max}  = 0.4$ to be consistent with the value of the jet resolution parameter used for the jet reconstruction.
The remaining contribution from the underlying event (UE) in the JC selection, which refers to particles not associated with jet fragmentation, is estimated in the perpendicular cone (PC) to the jet axis with radius $R = R_{\rm PC}$.
The default value is $R_{\rm PC} = 0.4$.

Since the $\eta$--$\varphi$ acceptance of the JC-selected particles differs from that for UE estimations, to subtract the UE component from the JC selection, a density distribution is defined
\begin{equation}\label{eq:normalize}
\frac{\dd\rho}{\dd\pT} = \frac{1}{N_{\rm ev}}\times\frac{1}{A_{\rm acc}}\times\frac{\dd N}{\dd\pT},
\end{equation}
where $\dd N/\dd\pT$ is the $\pT$-differential particle production yield, and $N_{\rm ev}$ and $A_{\rm acc}$ are the number of events and the area of the $\eta$--$\varphi$ acceptance for a given selection.
For JC and PC selections, $N_{\rm ev}$ corresponds to the number of events containing at least one jet with $\pTj^{\rm ch} > 10$~\GeVc.
The $\eta$--$\varphi$ acceptance area, $A_{\rm acc}$, is calculated via
\begin{equation}
A_{\rm acc} = \alpha\pi R^{2},
\end{equation}
where $R$ is the cone radius for the corresponding selection and $\alpha$ is a correction factor used to account for the partial geometrical overlap among jets on the $\eta$--$\varphi$ plane.
The $\alpha$ factor is calculated via a Monte Carlo~(MC) sampling approach using measured distributions of strange particles and jets as inputs.
The value of $\alpha$ is around $1.06$ and it is insensitive to particle species and event multiplicities.
It gives a minor correction on the particle density normalization since the production rate for jets with $\pTj^{\rm ch} > 10$~\GeVc is low even in high-multiplicity \pPb collisions.
With the normalization defined in Eq.~(\ref{eq:normalize}), the $\pT$-differential production density distribution of JE particles is obtained by subtracting the density distribution of particles with the UE selection from that with the JC selection, namely:
\begin{equation}
\label{eq:defJE}
\frac{\dd\rho^{\rm JE}}{\dd\pT} = \frac{\dd\rho^{\rm JC}}{\dd\pT} - \frac{\dd\rho^{\rm UE}}{\dd\pT}.
\end{equation}

In addition, to compare with the JE particles, the production yield of inclusive particles is also normalized according to Eq.~(\ref{eq:normalize}).
For the MB analysis, $N_{\rm ev}$ corresponds to the number of selected MB events.
For the event-activity differential analysis in \pPb collisions, $N_{\rm ev}$ corresponds to the number of selected events in the corresponding event activity interval.
The acceptance area, $A_{\rm acc}$, of particles in the inclusive analysis is given by
\begin{equation}
A_{\rm acc} = \Delta\eta\times\Delta\varphi,
\end{equation}
where $\Delta\eta = 2\times 0.75$ and $\Delta\varphi = 2\pi$, correspond to the $\eta$ and $\varphi$ acceptances of inclusive particles, respectively.

\begin{figure}[!t]
\begin{center}
\includegraphics[width=.49\textwidth]{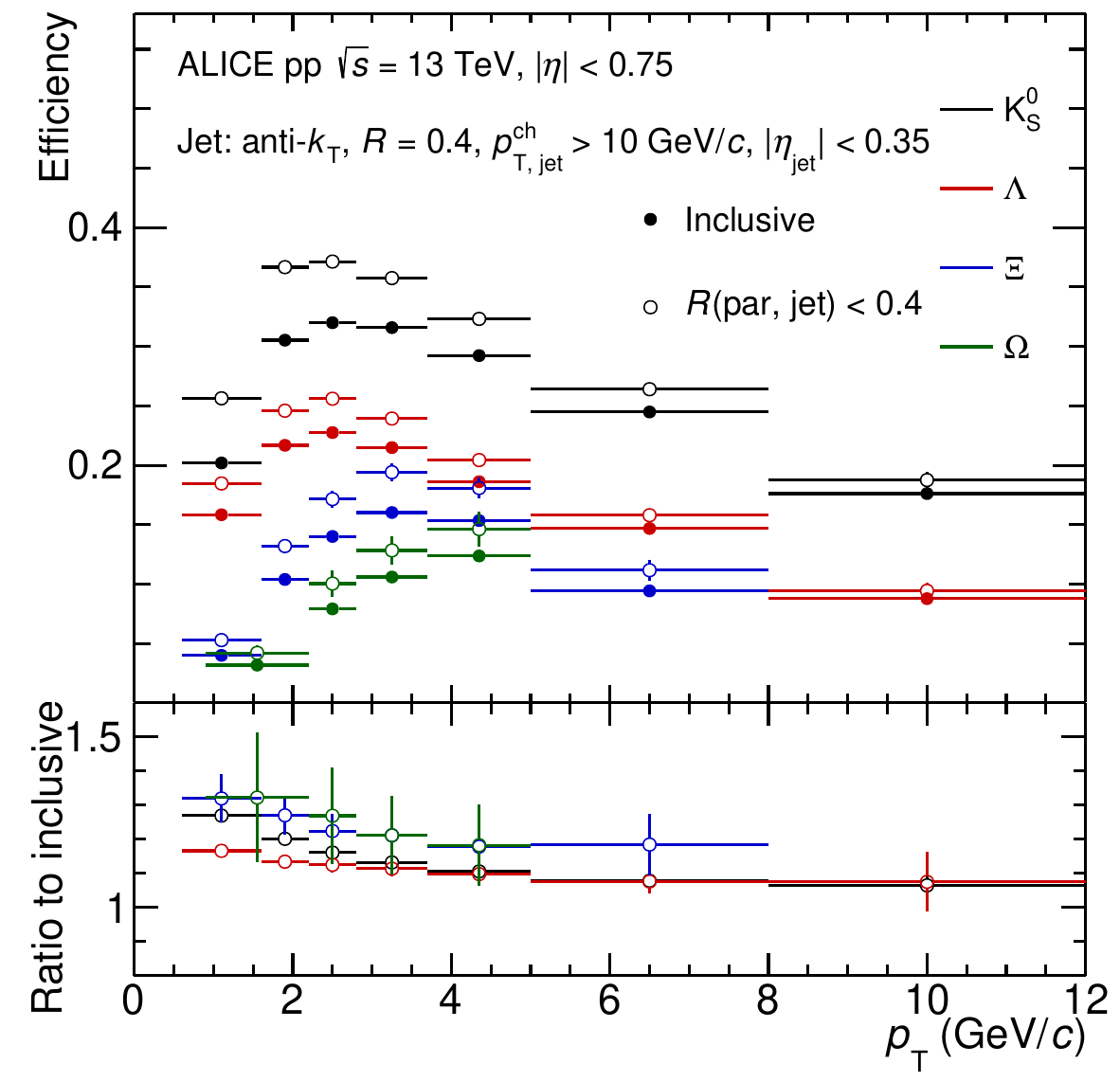}
\includegraphics[width=.49\textwidth]{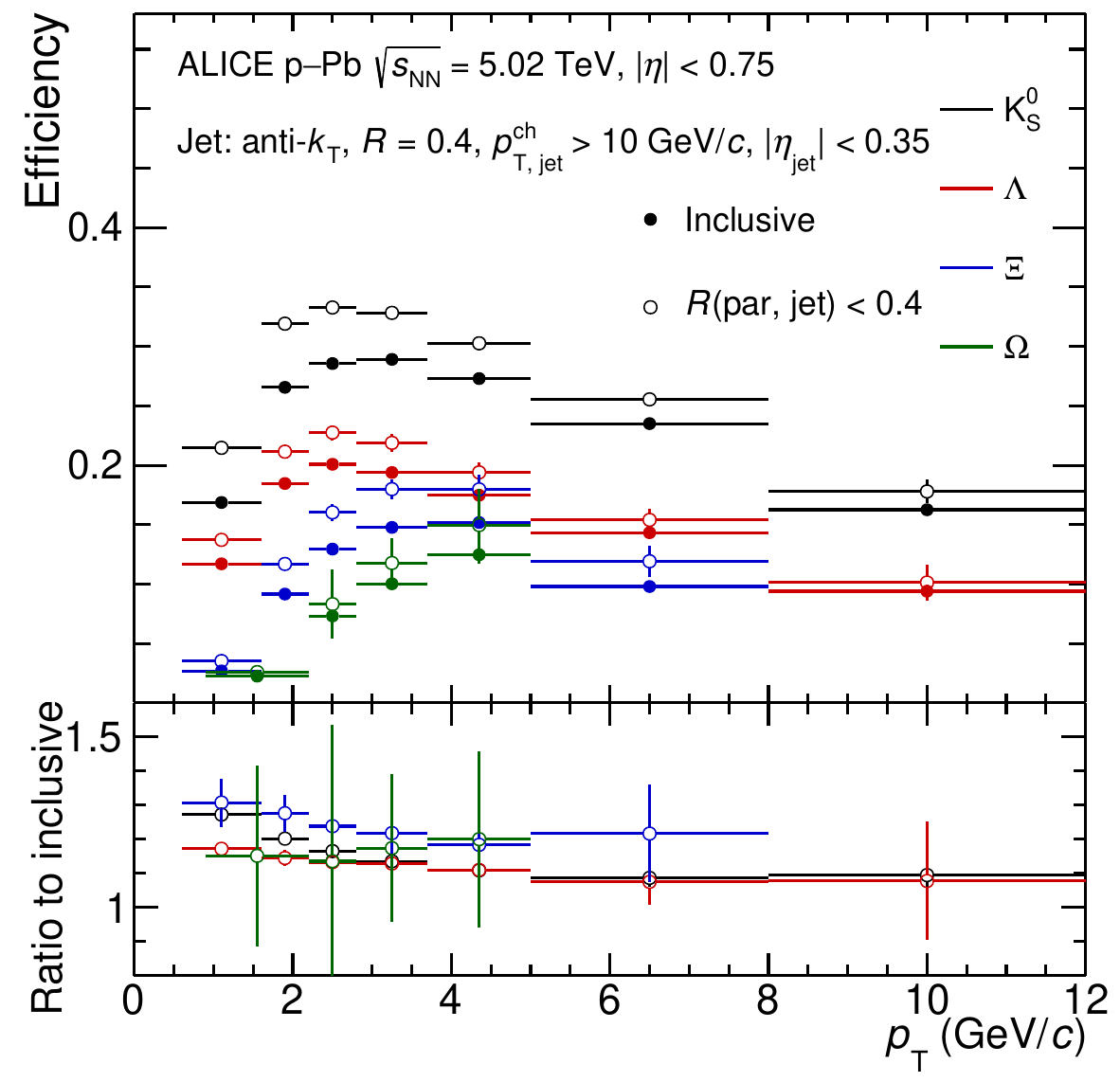}
\end{center}
\caption{Strange particle reconstruction efficiency in \pp collisions at \thirteen (left) and in \pPb collisions at \fivenn (right) for two selections: inside the jet cone ($R({\rm par, jet}) < 0.4$) and for the inclusive case.
The vertical bars represent the statistical uncertainties.}
\label{fig:EffiJCIncl}
\end{figure}

\subsection{Corrections for strange particle reconstruction and feed-down}
\label{SubSec:Correction}
The reconstruction efficiencies of each particle are obtained from MC simulated data.
For this purpose, $\Pyeight$ (for pp collisions)~\cite{Sjostrand:2014zea} and DPMJet (for \pPb collisions)~\cite{Roesler:2000he} are used and the simulated data are propagated through the ALICE detector by GEANT3~\cite{Brun:1994aa}.
Due to differences in the experimental acceptance for particles associated with jets and the underlying event, the efficiencies of particles are estimated separately for each case~\cite{Acharya:2021oaa}.
Figure~\ref{fig:EffiJCIncl} shows the reconstruction efficiency of the JC-selected particle and the inclusive one.
The reconstruction efficiencies of inclusive strange particles and those produced in the underlying event are identical within uncertainties.
For the JC-selected particles, the reconstruction efficiency is higher than that of inclusive particles at $\pT < 2$~\GeVc.
This is because the pseudorapidity distribution of strange particles matched with jets is narrower than that of inclusive ones and the $\eta$-differential reconstruction efficiency of strange particles decreases with $\abs{\eta}$.
This results in a higher $\eta$-integrated reconstruction efficiency of JC-selected particles than the inclusive ones.
This effect is more pronounced at low $\pT$.

\begin{figure}[!t]
\begin{center}
\includegraphics[width=.49\textwidth]{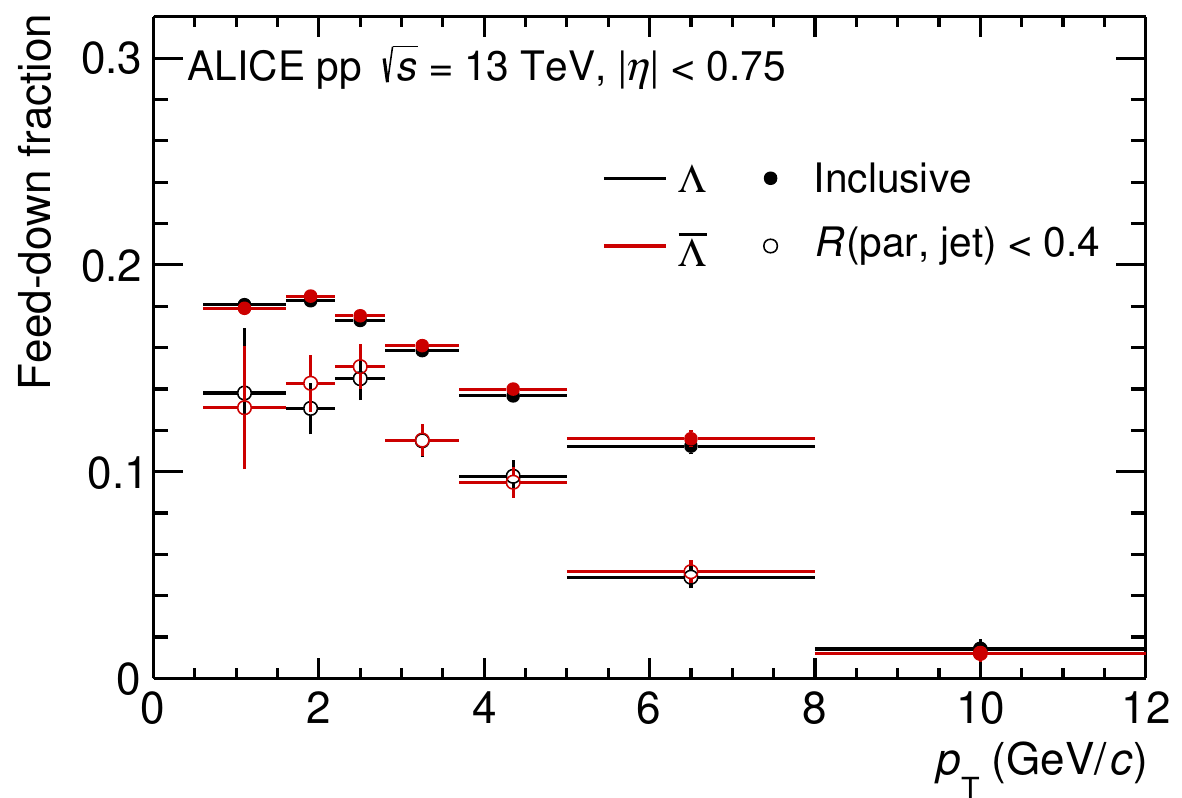}
\includegraphics[width=.49\textwidth]{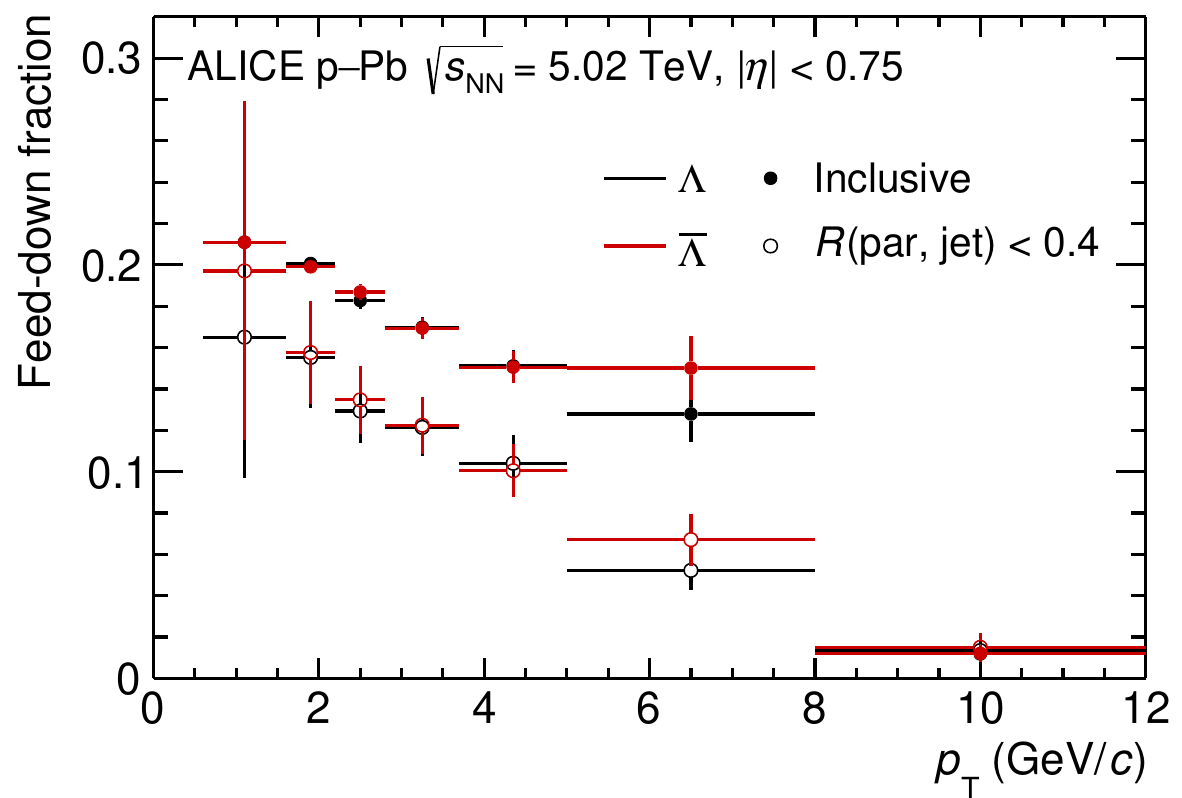}
\end{center}
\caption{Fraction of $\lmb$ yield removed due to the subtraction of feed-down contributions from charged and neutral $\Xi$ baryons decays in pp collisions at \thirteen (left) and \pPb collisions at \fivenn (right).
The vertical bars represent the statistical uncertainties.}
\label{fig:FdFrac}
\end{figure}

The yields for $\lmb$ and $\almb$ are significantly affected by secondary particles coming from the decays of charged and neutral $\Xi$ baryons.
The feed-down fraction is calculated with a data-driven approach~\cite{ALICE:2013wgn}.
The inclusive feed-down method was introduced in previous ALICE analyses~\cite{ALICE:2019avo,ALICE:2020jsh,ALICE:2017jyt}.
In this work, the feed-down fraction in jet and UE is computed for each $\pT$ interval using the measured spectra of $\Xi^\pm$ baryons in jet and UE.
The correction of the feed-down contribution from neutral $\Xi$ baryons is based on the assumption that the production rates of charged and neutral $\Xi$ baryons are equal.
Figure~\ref{fig:FdFrac} shows the results of the feed-down fraction for the JC selection compared with the inclusive one.

\subsection{Systematic uncertainties}%
\label{sec:SysUncer}

The total systematic uncertainties for $\kzero$, $\lmb$, $\almb$, $\Xis$, and $\Oms$ yields were estimated separately for each particle and in different $\pT$ intervals.
Individual selection criteria are loosened and tightened, in order to estimate the uncertainty on the discrepancy between data and MC simulations.
The main sources of the systematic uncertainty investigated in this measurement are related to the knowledge of detector materials, track selections, particle identification, proper lifetime, topological selections and signal extraction.
All individual uncertainty contributions are listed in Tables~\ref{tab:ppInclUncer} and~\ref{tab:pPbInclUncer} and finally added in quadrature.

\begin{table}[!t]
\begin{center}
\caption{Main sources and corresponding relative systematic uncertainties (in \%) for $\kzero$, $\lmb + \almb$, $\X + \Ix$, and $\Om + \Mo$ in \pp collisions at \thirteen.
The values are reported for low, intermediate, and high $\pT$.}
\label{tab:ppInclUncer}
\begin{tabularx}{\textwidth}{@{} lCCCCCCCCCCCC @{}}
\toprule
\textbf{Uncertainty source} & \multicolumn{3}{c}{\textbf{$\kzero$}}
                            & \multicolumn{3}{c}{\textbf{$\lmb + \almb$}}
                            & \multicolumn{3}{c}{\textbf{$\X + \Ix$}}
                            & \multicolumn{3}{c}{\textbf{$\Om + \Mo$}} \\
\cmidrule(lr){2-4} \cmidrule(lr){5-7} \cmidrule(lr){8-10} \cmidrule(lr){11-13}
$\pT$ (\GeVc) & 0.6 & 2 & 10 & 0.6 & 2 & 10 & 0.6 & 2 & 7 & 1 & 2 & 5 \\
\midrule
Detector material   & 4.0 & 4.0 & 4.0 & 4.0   & 4.0 & 4.0 & 4.0 & 4.0 & 4.0   & 4.0   & 4.0 & 4.0 \\
Competing rejection & 0.2 & 0.1 & 0.2 & negl. & 0.1 & 3.1 & -   & -   & -     & -     & -   & - \\
Track selection     & 1.5 & 1.2 & 0.4 & 0.6   & 1.4 & 1.3 & 2.8 & 0.1 & negl. & negl. & 1.5 & 0.2 \\
Particle identification & 0.1   & 0.1 & 0.1   & 0.3 & 0.2 & 1.1   & 1.9 & 1.7 & 2.4 & 3.9 & 8.7 & 6\\
Proper decay length         & negl. & 0.1 & negl. & 2.1 & 0.4 & negl. & -   & -   & -   & -   & -   & -\\
Topological       & 0.2 & 1.4 & negl. & 3.9 & 0.8 & 3.9 & 0.6 & 0.9 & 1.0 & 2.8 & 5.4 & 2.4 \\
Signal extraction & 0.8 & 1.1 & 1.1   & 0.3 & 0.5 & 1.7 & 3.0 & 1.0 & 0.5 & 2.3 & 4.6 & 3.0 \\
\midrule
\textbf{Total uncertainty} & 4.4 & 4.6 & 4.2 & 6.1 & 4.4 & 6.7 & 6.1 & 4.5 & 4.8 & 6.7 & 12.0 & 8.2 \\
\bottomrule
\end{tabularx}
\end{center}
\end{table}

\begin{table}[!t]
\begin{center}
\caption{Main sources and corresponding relative systematic uncertainties (in \%) for $\kzero$, $\lmb + \almb$, $\X + \Ix$, and $\Om + \Mo$ in \pPb collisions at \fivenn.
The values are reported for low, intermediate, and high $\pT$.}
\label{tab:pPbInclUncer}
\begin{tabularx}{\textwidth}{@{} lCCCCCCCCCCCC @{}}
\toprule
\textbf{Uncertainty source} & \multicolumn{3}{c}{\textbf{\kzero}}
                            & \multicolumn{3}{c}{\textbf{$\lmb + \almb$}}
                            & \multicolumn{3}{c}{\textbf{$\X + \Ix$}}
                            & \multicolumn{3}{c}{\textbf{$\Om + \Mo$}} \\
\cmidrule(lr){2-4} \cmidrule(lr){5-7} \cmidrule(lr){8-10} \cmidrule(lr){11-13}
$\pT$ (\GeVc) & 0.6 & 2 & 10 & 0.6 & 2 & 10 & 0.6 & 2 & 7 & 1 & 2 & 5 \\
\midrule
Detector material   & 4.0 & 4.0 & 4.0 & 4.0 & 4.0   & 4.0 & 4.0   & 4.0   & 4.0   & 4.0 & 4.0 & 4.0 \\
Competing rejection & 0.2 & 0.3 & 0.5 & 0.1 & negl. & 5.1 & -     & -     & -     & -   & -   & - \\
Track selection     & 1.4 & 1.7 & 1.8 & 0.2 & 1.3   & 1.4 & negl. & negl. & negl. & 1.3 & 2.5 & negl.\\
Particle identification & 0.1   & 0.2   & 0.2   & 0.3 & 0.2 & 1.0   & 3.1 & 1.2 & negl. & 8.1  & 13.7 & 5.9 \\
Proper decay length & negl. & negl. & negl. & 1.6 & 0.3 & negl. & 0.6 & 0.4 & negl. & negl.&  3.3 & negl.\\
Topological       & 4.4 & 0.6 & 1.9 & 3.9 & 0.9 & 2.7 & 1.3 & negl. & 2.6 & 1.2   & 4.8 & 3.7 \\
Signal extraction & 0.3 & 2.6 & 1.7 & 0.6 & 0.5 & 2.6 & 5.1 & 0.9   & 2.6 & negl. & 5.2 & negl. \\
\midrule
\textbf{Total uncertainty} & 6.1 & 5.1 & 5.1 & 5.7 & 4.3 & 6.1 & 7.4 & 4.3 & 5.4 & 9.2 & 16.4 & 8.0 \\
\bottomrule
\end{tabularx}
\end{center}
\end{table}

\textbf{Material budget.} The effect of the incomplete knowledge of the detector material budget is evaluated by comparing different MC simulations in which the material budget was increased and decreased by $4.5\%$.
This value corresponds to the uncertainty on the determination of the material budget by measuring photon conversions~\cite{Abelev:2014ffa}.
This particular systematic uncertainty is about $4\%$~\cite{ALICE:2020jsh}.

\textbf{Rejection of competing decays.}
A \kzero candidate is rejected if its invariant mass under the hypothesis of a \lmb or \almb lies in the window of $\pm 5$~\MeVmass around the mass of the \lmb or \almb, and a \lmb (\almb)  candidate is rejected if its invariant mass under the \kzero hypothesis lies in the window of $\pm 10$~\MeVmass around the \kzero mass.
To compute the uncertainty due to the competing selection, the analysis is redone with this rejection at $5$~\MeVmass to $3$~\MeVmass and $6$~\MeVmass for the \kzero. In the case of the \lmb or \almb, the rejection is removed entirely.
It is worth noting that the complete removal of the rejection in the $\kzero$ analysis causes signal extraction to be unstable at intermediate $\pT$ ($1.5$ -- $2.5$~\GeVc) due to the high $\lmb/\kzero$ ratio in that range, which in turn creates a non-linear background.

\textbf{Track selection.}
To estimate the systematic uncertainty due to the track selection, the analysis is redone with an increased number of required clusters in the TPC from the default $70$ to $80$ clusters out of a maximum of $159$ clusters.

\textbf{Particle identification.}
The TPC $\dEdx$ selection is used to reduce the combinatorial background in the strange particle invariant mass distribution.
The number of standard deviations $\sigma$ in the identification of particles using the $\dEdx$ has been varied from $4\sigma$ to $6\sigma$.

\textbf{Proper decay length selection.}
The proper decay length is defined as $mLc/p$, where $m$ is the mass of the particles, $L$ is the decay length, and $p$ is the particle's momentum.
The selection on the $mLc/p$ is varied within $12$ to $40$~cm for $\kzero$, $20$ to $40$ cm for $\lmb$ $(\almb)$, $10$ to $30$ cm for $\Xis$, and $5$ to $15$ cm for $\Oms$.

\textbf{Topological selection.}
The values of the selection criteria for the topological variables are varied within ranges resulting in a maximum variation of $\pm 10\%$ in the raw signal yield around its nominal value.
The observed deviations for each component are summed in quadrature.

\textbf{Signal extraction.}
In the same manner as for the topological selection, the signal extraction has been tested by varying the widths used to define the ``signal'' and ``background'' regions, expressed in terms of the number of $\sigma$ as defined in Section~\ref{sec:ParRec}.
In particular, the width of the peak region is varied from the default value of $6\sigma$ to $7\sigma$, $5\sigma$, and $4\sigma$ for \Vzero particles and  $3\sigma$ to $4\sigma~(3.5\sigma)$ and $2.5\sigma$ for $\Xi~(\Omega)$.

Additional systematic uncertainties on the particle yield originate from the UE subtraction and the jet $\pT$ threshold.
The systematic uncertainty due to the UE subtraction is estimated by varying the perpendicular cone radius from the chosen default value of $R_{\rm PC} = 0.4$ to $0.2$ and $0.6$.
From the deviations obtained for different sizes of the PC, the relative systematic uncertainty of the UE subtraction is estimated.
To evaluate the uncertainty related to the jet $\pT$ threshold, the analysis is repeated varying the jet $\pT$ threshold by $\pm 1$~\GeVc.
The systematic uncertainties of particles in jets are added to the list of uncertainties in quadrature.
The values are shown in Table~\ref{tab:ppJEUncer} and Table~\ref{tab:pPbJEUncer}.

The uncertainties of JE-particle yield ratios ($\lmb/\kzero$, $\Xis/\kzero$, $\Oms/\kzero$, $\Xis/\lmb$, $\Oms/\lmb$, and $\Oms/\Xis$) include three sources: the particle reconstruction, UE subtraction, and the jet $\pT$ threshold.
The uncertainty on particle reconstruction is propagated from that obtained for particle spectra.
The systematic uncertainties related to the material budget are correlated for each particle spectrum and they partially cancel in the ratios.
Uncertainties related to UE subtraction and jet $\pT$ threshold are obtained by varying the same condition for particle spectra in both numerator and denominator of the corresponding ratios.
For JE particles, the uncertainty on UE subtraction is only significant at low $\pT$ since the underlying event is dominant in this region.
In contrast, the uncertainty on the variation of the jet $\pT$ threshold is more pronounced at high $\pT$ because the $\pT$ slope of particles produced in jets is sensitive to the jet $\pT$ threshold.

\begin{table}[!t]
\begin{center}
\caption{Main sources and corresponding relative systematic uncertainties (in \%) for particle $\pT$-differential density ($\kzero$, $\lmb + \almb$, $\X + \Ix$, and $\Om + \Mo$) and particle ratios ($\lmb/\kzero$, $\Xi/\kzero$, $\Omega/\kzero$, $\Xi/\lmb$, $\Omega/\lmb$, and $\Omega/\Xi$) in JE in \pp collisions at \thirteen.
The values are reported for low (0.6~\GeVc), intermediate (2~\GeVc), and high (10~\GeVc) $\pT$.}
\label{tab:ppJEUncer}
\begin{tabularx}{\textwidth}{@{} lCCCCCCCCCCCC @{}}
\toprule
\textbf{Uncertainty source} & \multicolumn{3}{c}{\textbf{\kzero}}
                            & \multicolumn{3}{c}{\textbf{$\lmb + \almb$}}
                            & \multicolumn{3}{c}{\textbf{$\X + \Ix$}}
                            & \multicolumn{3}{c}{\textbf{$\Om + \Mo$}} \\
\cmidrule(lr){2-4} \cmidrule(lr){5-7} \cmidrule(lr){8-10} \cmidrule(lr){11-13}
$\pT$ (\GeVc) & 0.6 & 2 & 10 & 0.6 & 2 & 10 & 0.6 & 2 & 7 & 1 & 2 & 5 \\
\midrule
Particle reconstruction & 1.8 & 0.3 & negl. & 5.5 & 0.6 & negl. & 6.7 & 0.9 & 0.1 & 6.0   & 1.7   & 0.3\\
UE subtraction          & 0.1 & 0.1 &  0.1  & 0.1 & 0.2 & 0.1   & 1.5 & 0.2 & 0.3 & 3.6   & 1.8   & 0.5\\
Jet $\pT$ threshold     & 0.6 & 3.1 & 10.9  & 0.6 & 1.1 & 9.9   & 3.5 & 2.4 & 5.0 & negl. & negl. & negl. \\
\midrule
\textbf{Total uncertainty}& 1.8& 3.1& 10.9& 5.6& 1.2& 9.9& 7.7& 2.6& 5& 7.1& 2.5& 0.6 \\
\bottomrule
\end{tabularx}

\begin{tabularx}{\textwidth}{@{} lCCCCCCCCC @{}}
\toprule
\textbf{Uncertainty source} & \multicolumn{3}{c}{\textbf{$(\lmb + \almb)/(2\kzero)$}}
                            & \multicolumn{3}{c}{\textbf{$(\X + \Ix)/(2\kzero)$}}
                            & \multicolumn{3}{c}{\textbf{$(\Om + \Mo)/(2\kzero)$}} \\
\cmidrule(lr){2-4} \cmidrule(lr){5-7} \cmidrule(lr){8-10}
$\pT$ (\GeVc) & 0.6 & 2 & 10  & 0.6 & 2 & 7 & 1 & 2 & 5 \\
\midrule
Particle reconstruction & 2.4 & 2.8 & 3.3 & 3.4 & 2.8 & 2.8 &  6.7 & 11.4 & 7.3 \\
UE subtraction          & 0.8 & 0.2 & 0.4 & 3.5 & 0.2 & 0.1 & 10.0 &  4.0 & 2.2 \\
Jet $\pT$ threshold     & 0.4 & 2.3 & 1.0 & 1.7 & 1.6 & 3.6 &  1.0 &  3.3 & 6.4 \\
\midrule
\textbf{Total uncertainty} & 2.6 & 3.7 & 3.5 & 5.2 & 3.3& 4.5 & 12.4 & 12.5 & 10.0 \\
\bottomrule
\end{tabularx}

\begin{tabularx}{\textwidth}{@{} lCCCCCCCCC @{}}
\toprule
\textbf{Uncertainty source} & \multicolumn{3}{c}{\textbf{$(\X + \Ix)/(\lmb + \almb)$}}
                            & \multicolumn{3}{c}{\textbf{$(\Om + \Mo)/(\lmb + \almb)$}}
                            & \multicolumn{3}{c}{\textbf{$(\Om + \Mo)/(\X + \Ix)$}} \\
\cmidrule(lr){2-4} \cmidrule(lr){5-7} \cmidrule(lr){8-10}
$\pT$ (\GeVc) & 0.6 & 2 & 7 & 1 & 2 & 5 & 1 & 2 & 5 \\
\midrule
Particle reconstruction & 3.4 & 3.0 & 3.2 &  6.7 & 11.5 & 7.5 & 6.8 & 11.5 & 7.4 \\
UE subtraction          & 4.4 & 0.4 & 0.2 & 12.4 &  4.2 & 2.3 & 7.8 &  3.8 & 2.7 \\
Jet $\pT$ threshold     & 0.7 & 0.5 & 1.9 &  0.2 &  0.9 & 3.5 & 0.4 &  1.3 & 3.0 \\
\midrule
\textbf{Total uncertainty} & 5.6 & 3.0 & 3.7 & 14.1 & 12.2 & 8.6 & 10.3 & 12.1 & 8.5 \\
\bottomrule
\end{tabularx}
\end{center}
\end{table}

\begin{table}[!t]
\begin{center}
\caption{Main sources and corresponding relative systematic uncertainties (in \%) for particle $\pT$-differential density ($\kzero$, $\lmb + \almb$, $\X + \Ix$, and $\Om + \Mo$) and particle yield ratios ($\lmb/\kzero$, $\Xi/\kzero$, $\Omega/\kzero$, $\Xi/\lmb$, $\Omega/\lmb$, and $\Omega/\Xi$) in JE in \pPb collisions at \fivenn.
The values are reported for low, intermediate, and high $\pT$.}
\label{tab:pPbJEUncer}
\begin{tabularx}{\textwidth}{@{} lCCCCCCCCCCCC @{}}
\toprule
\textbf{Uncertainty source} & \multicolumn{3}{c}{\textbf{\kzero}}
                            & \multicolumn{3}{c}{\textbf{$\lmb + \almb$}}
                            & \multicolumn{3}{c}{\textbf{$\X + \Ix$}}
                            & \multicolumn{3}{c}{\textbf{$\Om + \Mo$}} \\
\cmidrule(lr){2-4} \cmidrule(lr){5-7} \cmidrule(lr){8-10} \cmidrule(lr){11-13}
$\pT$ (\GeVc) & 0.6 & 2 & 10 & 0.6 & 2 & 10 & 0.6 & 2 & 7 & 1 & 2 & 5 \\
\midrule
Particle reconstruction & 5.0 & 0.8 & negl. & 14.2  & 1.5 & negl. & 24.8 & 2.8 & 0.3 & 8.7   & 3.7   & 0.9 \\
UE subtraction          & 0.3 & 0.1 & 0.1   & negl. & 0.1 & 11.2  & 14.1 & 0.8 & 0.7 & negl. & negl. & 1.2 \\
Jet $\pT$ threshold     & 0.3 & 3.5 & 11    & 3.2   & 1.8 & 0.1   & 24.9 & 3.0 & 4.1 & 3.1   & 10.7  & 7.6 \\
\midrule
\textbf{Total uncertainty} & 5.0 & 3.6 & 11.0 & 14.6 & 2.3 & 11.2 & 37.9 & 4.2 & 4.1 & 9.3 & 11.3 & 7.7 \\
\bottomrule
\end{tabularx}

\begin{tabularx}{\textwidth}{@{} lCCCCCCCCC @{}}
\toprule
\textbf{Uncertainty source} & \multicolumn{3}{c}{\textbf{$(\lmb + \almb)/(2\kzero)$}}
                            & \multicolumn{3}{c}{\textbf{$(\X + \Ix)/(2\kzero)$}}
                            & \multicolumn{3}{c}{\textbf{$(\Om + \Mo)/(2\kzero)$}} \\
\cmidrule(lr){2-4} \cmidrule(lr){5-7} \cmidrule(lr){8-10}
$\pT$ (\GeVc) & 0.6 & 2 & 10  & 0.6 & 2 & 7 & 1 & 2 & 5 \\
\midrule
Particle reconstruction & 3.3 & 3.4 & 4.7 & 4.7 & 3.2 & 4.0 & 9.8 & 1.5   & 7.4 \\
UE subtraction          & 0.8 & 0.1 & 0.1 & 9.1 & 1.8 & 1.0 & 4.1 & negl. & 0.3 \\
Jet $\pT$ threshold     & 1.4 & 2.6 & 0.1 & 8.6 & 2.4 & 6.0 & 0.5 & 1.5   & 0.3 \\
\midrule
\textbf{Total uncertainty} & 3.7 & 4.3 & 4.7 & 13.4 & 4.4 & 7.2 & 10.6 & 15.1 & 7.4 \\
\bottomrule
\end{tabularx}

\begin{tabularx}{\textwidth}{@{} lCCCCCCCCC @{}}
\toprule
\textbf{Uncertainty source} & \multicolumn{3}{c}{\textbf{$(\X + \Ix)/(\lmb + \almb)$}}
                            & \multicolumn{3}{c}{\textbf{$(\Om + \Mo)/(\lmb + \almb)$}}
                            & \multicolumn{3}{c}{\textbf{$(\Om + \Mo)/(\X + \Ix)$}} \\
\cmidrule(lr){2-4} \cmidrule(lr){5-7} \cmidrule(lr){8-10}
$\pT$ (\GeVc) & 0.6 & 2 & 10  & 0.6 & 2 & 7 & 1 & 2 & 5 \\
\midrule
Particle reconstruction & 4.3 & 2.8 & 3.8 & 9.6 & 15.0 & 7.5 & 10.0 & 14.9 & 8.6 \\
UE subtraction          & 9.9 & 1.9 & 0.8 & 3.6 &  0.1 & 0.3 & 11.0 &  1.8 & 0.1 \\
Jet $\pT$ threshold     & 2.7 & 0.6 & 2.6 & 0.4 &  5.0 & 3.0 &  0.4 &  3.8 & 1.7 \\
\midrule
\textbf{Total uncertainty} & 11.1 & 3.4 & 4.7 & 10.3 & 15.8 & 8.1 & 14.9 & 15.5 & 8.7 \\
\bottomrule
\end{tabularx}
\end{center}
\end{table}

\section{Results and discussion}%
\label{sec:Results}

\subsection{Particle production and yield ratios in pp collisions at \thirteen}

\begin{figure}[!tp]
\begin{center}
\includegraphics[width=.8\textwidth]{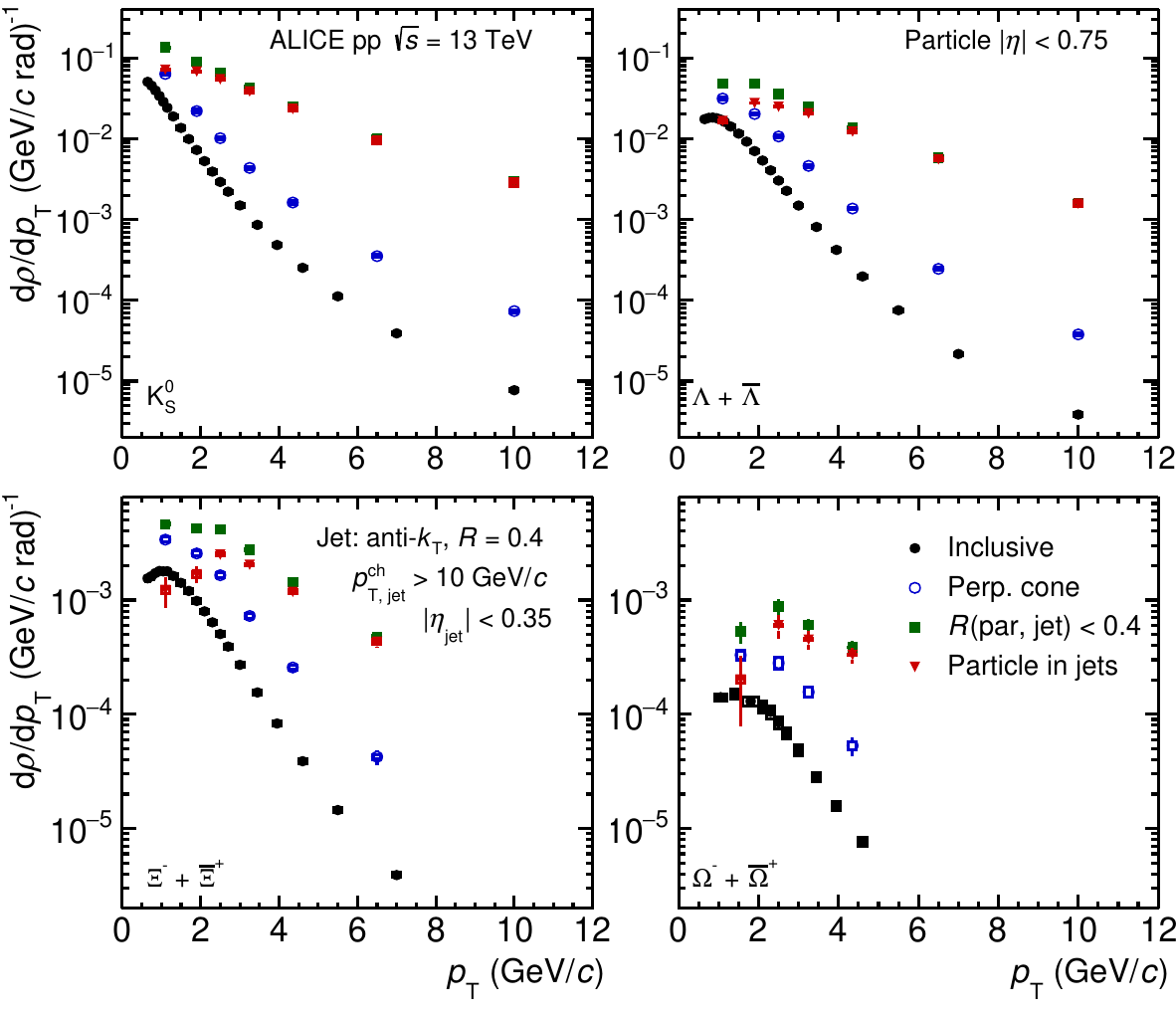}
\end{center}
\caption{$\pT$-differential density, $\dd\rho/\dd\pT$, of $\kzero$ (top left panel), $\lmb$ (top right panel), $\Xi$ (bottom left panel), and $\Omega$ (bottom right panel) in pp collisions at \thirteen.
The spectra of JE particles (red triangles), associated with hard scatterings, are compared with that of JC (green squares) and UE (blue open circles) selections.
The results from inclusive measurements (black closed circles) are presented as well.
The statistical uncertainties are represented by the vertical error bars and the systematic uncertainties by the boxes.}
\label{fig:ppSpect}
\end{figure}

\begin{figure}[!tp]
\begin{center}
\includegraphics[width=1.\textwidth]{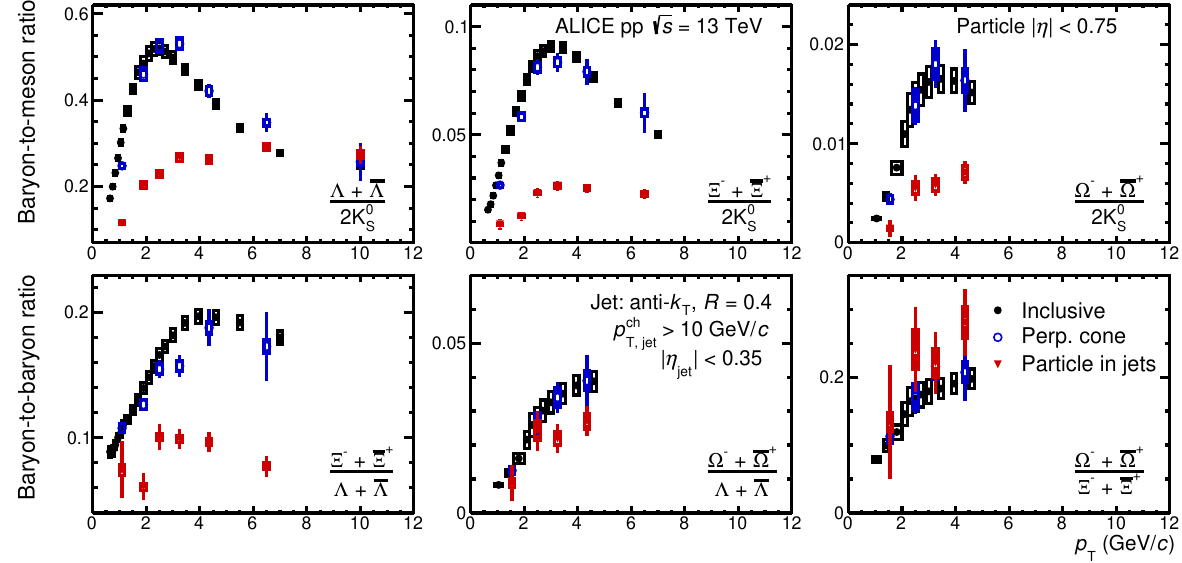}
\end{center}
\caption{$\pT$-dependent strange baryon-to-meson (top) and baryon-to-baryon (bottom) yield ratios in pp collisions at \thirteen.
For each case, the results of JE particles (red triangles) are compared with that of inclusive (black closed circles) and UE (blue open circles) particles.
The statistical uncertainties are represented by the vertical error bars and the systematic uncertainties by the boxes.}
\label{fig:ppRdata}
\end{figure}

For the strange hadrons discussed in this paper, the ratios of yields for particles and antiparticles are consistent with unity within uncertainties, as expected at LHC energies in the midrapidity region~\cite{ALICE:2020jsh, ALICE:2019avo, ALICE:2015mpp}.
Therefore, all the spectra and the corresponding ratios shown in the following are reported after summing particles and antiparticles, when a distinct antiparticle state exists.
The sums of particles and antiparticles, $\lmb + \almb$, $\X + \Ix$, and $\Om + \Mo$ are simply denoted as $\lmb$, $\Xi$, and $\Omega$, unless explicitly written.

Figure~\ref{fig:ppSpect} shows the fully corrected $\pT$-differential densities, $\dd\rho/ \dd\pT$ defined by Eq.~(\ref{eq:normalize}), of \kzero, $\lmb$, $\Xi$, and $\Omega$ associated with charged-particle jets and the underlying event in \pp collisions at \thirteen.
For particles matched to the jet cone, the JC-selected particles defined by Eq.~(\ref{eq:defJC}), the UE component estimated using the PC selection is mainly concentrated at low $\pT$ in the region of $\pT< 1$--$2$~\GeVc.
The UE fraction is higher for $\lmb$, $\Xi$, and $\Omega$ baryons than for $\kzero$ mesons.
In the high-$\pT$ region ($\pT > 5$~\GeVc), JC-selected particle production is dominated by the products from hard scatterings -- the JE particles defined by Eq.~(\ref{eq:defJE}).
The density distribution of UE particles rapidly decreases with $\pT$, reaching values about one order of magnitude lower than that with the JC selection for particle $\pT$ exceeding $4$~\GeVc.
This is consistent with the expectation that the high-$\pT$ particles originate from jet fragmentation.
The density distributions of the inclusive particles from the inclusive analysis are also compared with that of the JE and UE particles in Fig.~\ref{fig:ppSpect}.
Since the density of JE particles is obtained from events triggered by charged-particle jets, its $\pT$-dependence is considerably less steep than that of inclusive particles.
The density of UE particles is higher than that of inclusive particles since the UE is obtained from events containing jets with $\pTjch > 10$~\GeVc.
This is due to the well established jet pedestal effect~\cite{ALICE:2019mmy}.
The underlying event density increases with the leading track $\pT$ and then saturates at high $\pT$.
The $\pT$ dependence of the inclusive particle density is qualitatively similar to that for UE particles, which shows a steeply falling distribution with increasing $\pT$.

\begin{figure}[!p]
\begin{center}
\includegraphics[width=.49\textwidth]{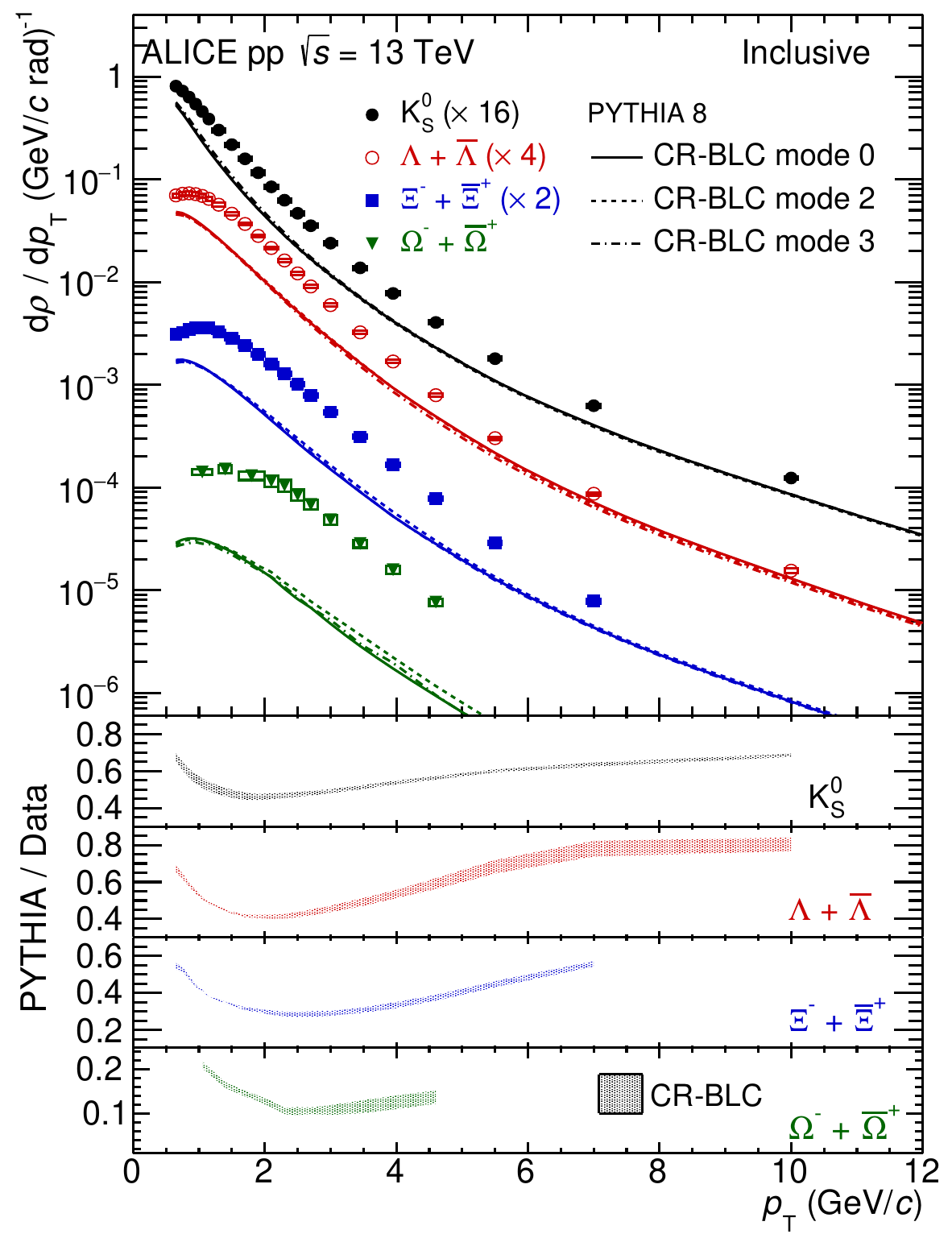}
\includegraphics[width=.49\textwidth]{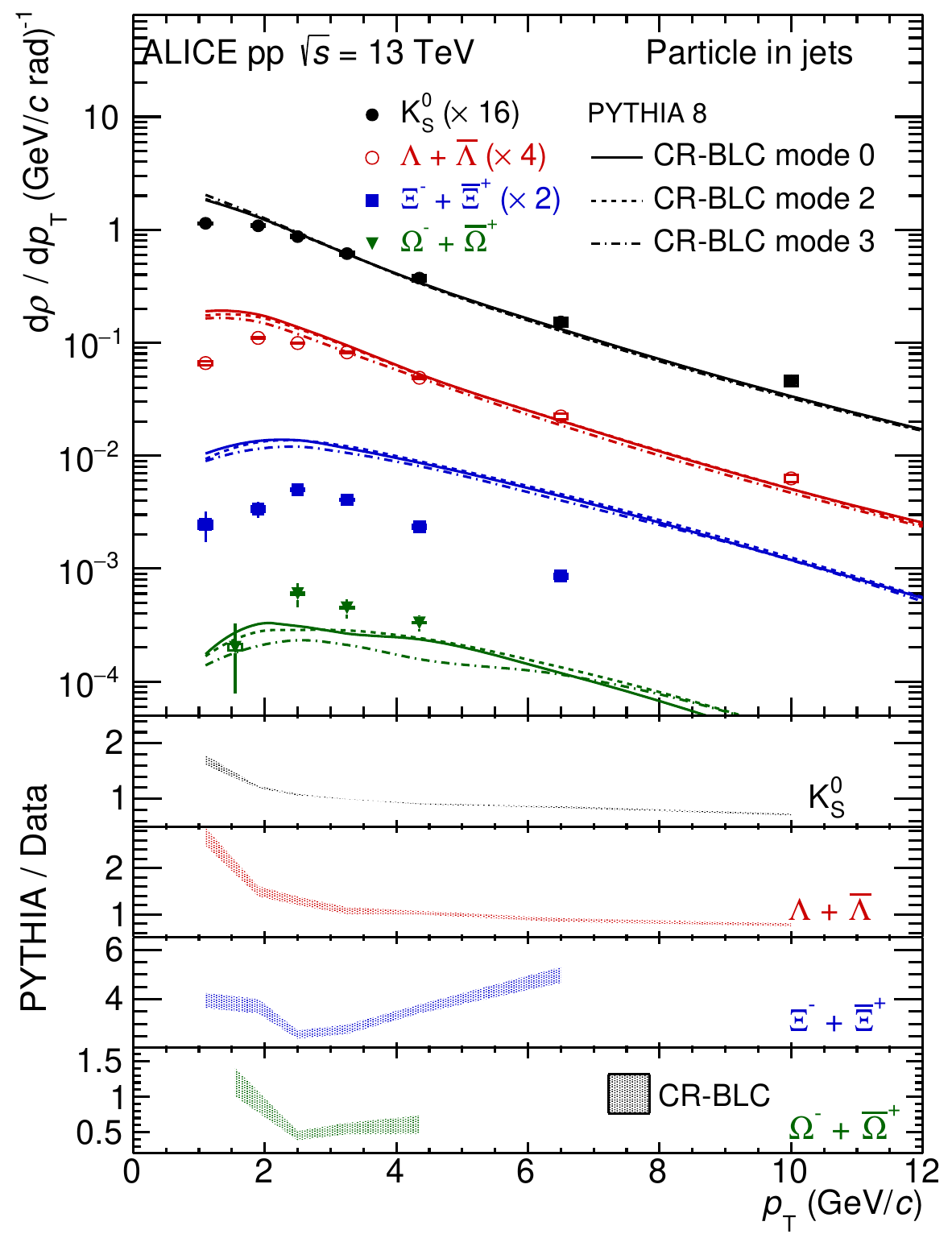}
\end{center}
\caption{$\pT$-differential density distributions for inclusive (left) and within jets (right) $\kzero$ (black closed circles), $\lmb$ (red open circles), $\Xi$ (blue squares), and $\Omega$ (green inverted triangles) in \pp collisions at \thirteen.
The spectra in data are compared with \Pyeight CR-BLC simulations.
Three modes, labeled as mode $0$ (solid line), $2$ (dashed line), and $3$ (dash-dotted line) are adopted in the simulations.
The \Pdratio ratios are shown in the four bottom panels where the spread of the three \Pyeight CR-BLC implementation modes are presented as bands.
For clarity, some of the spectra were scaled with the factors indicated in the legends.
The statistical uncertainties are represented by the vertical error bars and the systematic uncertainties by the boxes.}
\label{fig:ppIncandJE}
\end{figure}

The yield ratios of $\lmb$, $\Xi$, and $\Omega$ baryons to $\kzero$ mesons as a function of $\pT$ in \pp collisions at \thirteen are shown in the top three panels of Fig.~\ref{fig:ppRdata} for different selection criteria.
The yield ratios of the UE selected particles given by the PC selection are consistent with that of inclusive particles within uncertainties.
These ratios show an enhancement in the $\pT$ region $1$--$5$~\GeVc with respect to those for the JE particles.
The yield ratios of the JE selected particles are approximately independent of $\pT$ in the region beyond $2$~\GeVc, in particular, they do not show a maximum at intermediate $\pT$.
Clearly the enhancement of the baryon-to-meson yield ratio seen in the inclusive measurement is not present within jets.
It is worth noting that the $\lmb/\kzero$ ratio of inclusive particles becomes consistent with that of JE particles within uncertainties for $\pT > 6$~\GeVc as the high-$\pT$ inclusive particles originate predominantly from jet fragmentation. The results for JE particles are consistent with those in Ref.~\cite{Acharya:2021oaa} for pp collisions at \seven, showing no dependence on the collision energy.
For the $\Xi$/$\kzero$ and $\Omega$/$\kzero$ ratios, even with limited $\pT$ coverage, the trends imply that the results of inclusive measurements are expected to converge to that of JE particles at higher $\pT$.

The strange baryon-to-baryon yield ratios, $\Xi$/$\lmb$, $\Omega$/$\lmb$, and $\Omega$/$\Xi$, as a function of $\pT$ are presented in the bottom three panels of Fig.~\ref{fig:ppRdata}.
For each case, the numerator always contains at least one more strange quark than the denominator, and the results obtained from JE, UE and inclusive selections are compared.
Similarly to the baryon-to-meson yield ratios shown in the upper three panels of the same figure, the baryon-to-baryon yield ratios from the UE are consistent with that from the inclusive measurements within uncertainties.
In the measured $\pT$ range, they increase with $\pT$ up to $4$~\GeVc.
For $\pT > 2$~\GeVc, the $\Xi$/$\lmb$ ratio of JE particles is almost independent of $\pT$, and shows a strong suppression by a factor of about two with respect to the inclusive measurements.
However, the $\Omega$/$\lmb$ and $\Omega$/$\Xi$ ratios for particles associated with jets show a similar $\pT$ dependence as the inclusive particles.
The $\Omega$/$\lmb$ ratio of JE particles is systematically lower than the inclusive measurement, but with a smaller suppression than that observed for $\Xi/\lmb$.
For the $\Omega$/$\Xi$ ratio, the results of JE particles are compatible with that of inclusive particles within uncertainties.
These findings suggest that the production mechanism of $\Omega$ baryons, as strange-quark triplets, in jets may be similar to that in the UE.
This conclusion can be further confirmed in future measurements using a larger data sample.

The $\pT$-differential densities of inclusive $\kzero$, $\lmb$, $\Xi$, and $\Omega$ are compared with simulations with \Pyeight event generator~\cite{Sjostrand:2014zea} in the left panels of Fig.~\ref{fig:ppIncandJE}.
The \Pyeight Monte Carlo simulation studies are performed with the CR-BLC model~\cite{Christiansen:2015yqa}, in which the minimisation of the string potential is implemented considering the SU($3$) multiplet structure of QCD, allowing for the formation of ``baryonic'' configurations where two colours can combine coherently to form anti-colours.
From the CR-BLC model~\cite{Christiansen:2015yqa}, three modes (labeled as mode $0$, $2$, and $3$) are suggested by the authors -- each applying different constraints on the allowed reconnections among the colour sources.
In particular, considerations are given to the causal connections among strings involved in the reconnection and to the time dilation caused by relative boosts of the strings.
The density spectra using \Pyeight are normalized in the same way as data described in Section~\ref{sec:ParJetMatch}.
The left-bottom panels of Fig.~\ref{fig:ppIncandJE} show the ratios between \Pyeight simulations and data.
Since for each particle species the three CR-BLC modes give almost identical $\pT$-differential density spectra, the results corresponding to different CR-BLC modes are presented as bands (unless explicitly stated otherwise).

The inclusive density spectra obtained with \Pyeight underestimate the data for all particle species and the $\pT$ dependence follows a power-law trend, which does not reproduce the moderate peaks on the spectra around $\pT = 2$~\GeVc present in data.
This results in a ``valley'' structure in \Pdratio ratios in the interval of $1 < \pT < 4$~\GeVc.
For $\kzero$ and $\lmb$, the value of the MC/data ratio reaches the minimum of around $0.4$ at $\pT\simeq 2$~\GeVc, then it increases with $\pT$ for $\pT > 2$~\GeVc and shows a saturation trend with a value that rises to $0.8$ at $\pT > 6$~\GeVc.
For the multi-strange baryons, $\Xi$ and $\Omega$, the ratio decreases with strange-quark content and baryon mass.
The minimum values of the ratio are around $0.2$ and $0.1$ for $\Xi$ and $\Omega$, respectively.

In analogy with the left panel of Fig.~\ref{fig:ppIncandJE}, the right panel shows the comparisons of $\pT$-differential densities for JE particles with the corresponding \Pyeight simulations for the different CR-BLC modes.
The JE-particle density spectra from \Pyeight are obtained following the same approach applied to data as detailed in Section~\ref{sec:ParJetMatch}.
\Pyeight simulations overestimate the density of $\kzero$ mesons and $\lmb$ baryons in jets for $\pT < 2$~\GeVc, while, in general, a better agreement is observed for $\pT > 2$~\GeVc.
The MC/data ratio for those two particle species are almost identical, as seen in the lower panels of Fig.~\ref{fig:ppIncandJE}.
But, in general, the $\pT$-differential density obtained from the generator is softer than that in the data.
For $\Xi$ baryons, \Pyeight overestimates their production associated with jets over the measured $\pT$ range, $0.9 < \pT < 8$~\GeVc, by a factor of around three to six depending on $\pT$.
A possible explanation is that the strings containing partons produced in the hard scattering processes have higher string tension during the PYTHIA fragmentation.

It is likely that in \Pyeight the ss-diquark string production rate is much higher than that within the jet fragmentation found in data.
Moreover, since the probability for an ss-diquark combining with another single s-quark to form an $\Omega$ baryon is lower than the probability combining with u- and d-quarks to form a $\Xi$ baryon, the density of $\Omega$ produced in jets is underestimated.
For $\pT > 2$~\GeVc, the corresponding $\Omega$ MC/data ratio reaches about $0.5$ with mild dependence on $\pT$.
This is indicative of the vastly overestimated production density of $\Xi$ baryons in jets within the generator.

\begin{figure}[!t]
\begin{center}
\includegraphics[width=1.\textwidth]{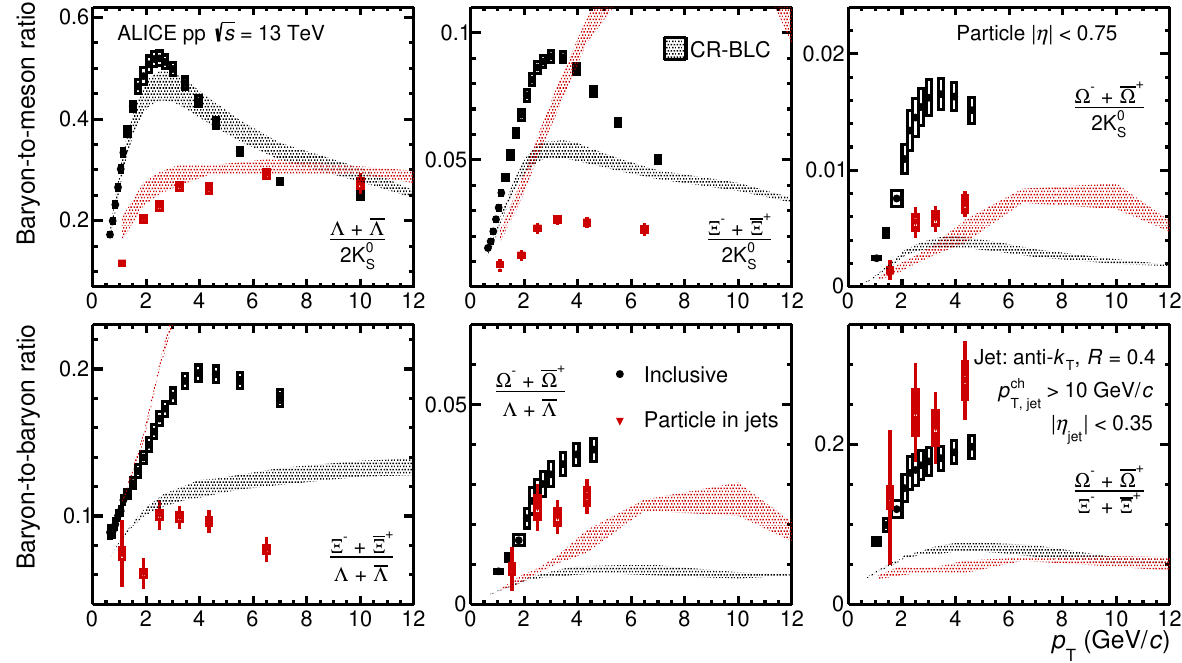}
\end{center}
\caption{$\pT$-dependent strange baryon-to-meson (top) and baryon-to-baryon (bottom) ratios in pp collisions at \thirteen.
For each case, the results of JE (red triangles) and inclusive (black closed circles) particles are compared with \Pyeight CR-BLC simulations.
The bands correspond to the spread of simulations of the three different CR-BLC implementation modes.
The statistical uncertainties are represented by the vertical error bars and the systematic uncertainties by the boxes.}
\label{fig:ppRmodel}
\end{figure}

The $\pT$-differential particle ratios from the JE and the inclusive selections are compared with the \Pyeight simulations in Fig.~\ref{fig:ppRmodel}.
\Pyeight CR-BLC tunes generally agree with the $\lmb$/$\kzero$ ratios for both JE particles and inclusive measurements, despite that the simulations do not reproduce the individual density spectra neither in jets nor in the inclusive sample.
Large discrepancies between data and the MC model are observed for all the other cases containing multi-strange hadrons in the numerator over the measured $\pT$ acceptance.
As stated in Ref.~\cite{Bierlich:2015rha}, although the string junction mechanism applied in \Pyeight CR-BLC tunes increases the baryon production probabilities, the ss-diquark is disfavoured in the \Pythia fragmentation due to the phase-space constraint on high invariant mass strings.
This results in \Pyeight largely underestimating the inclusive particle ratios containing multi-strange hadrons in the numerator.
Since the density of $\Xi$ in jets is overestimated by \Pyeight, then the $\Xi$/$\kzero$ and $\Xi$/$\lmb$ ratios given by \Pyeight increase dramatically with $\pT$ and raise to unrealistic large values.
At the same time, the $\Omega$/$\Xi$ ratio in jets is suppressed in \Pyeight generated events.
It seems that \Pyeight qualitatively reproduces the $\pT$ dependence for $\Omega$/$\kzero$ and $\Omega$/$\lmb$ ratios in jets.
However, this may be due to the unrealistic enhancement of the ss-diquark produced by strings containing partons from hard scatterings in the model.
It is worth noticing that the colour rope mechanism~\cite{Bierlich:2014xba}, in which the strange particle production is enhanced via interactions between strings~\cite{Bierlich:2015rha}, vastly overestimates the $\Xi$ and $\Omega$ production in jets at high $\pT$ (see the illustration shown in Fig.~\ref{fig:Rope} in Appendix~\ref{app:ppRope}).
The enhancement of the multi-strange production in colour rope predictions results from the higher local energy density in the region where an energetic jet is present.
In summary, the measurements presented in this section provide important constraints on the production mechanisms of particles, especially for the multi-strange baryon, associated with hard partons.

\begin{figure}[!t]
\begin{center}
\includegraphics[width=.8\textwidth]{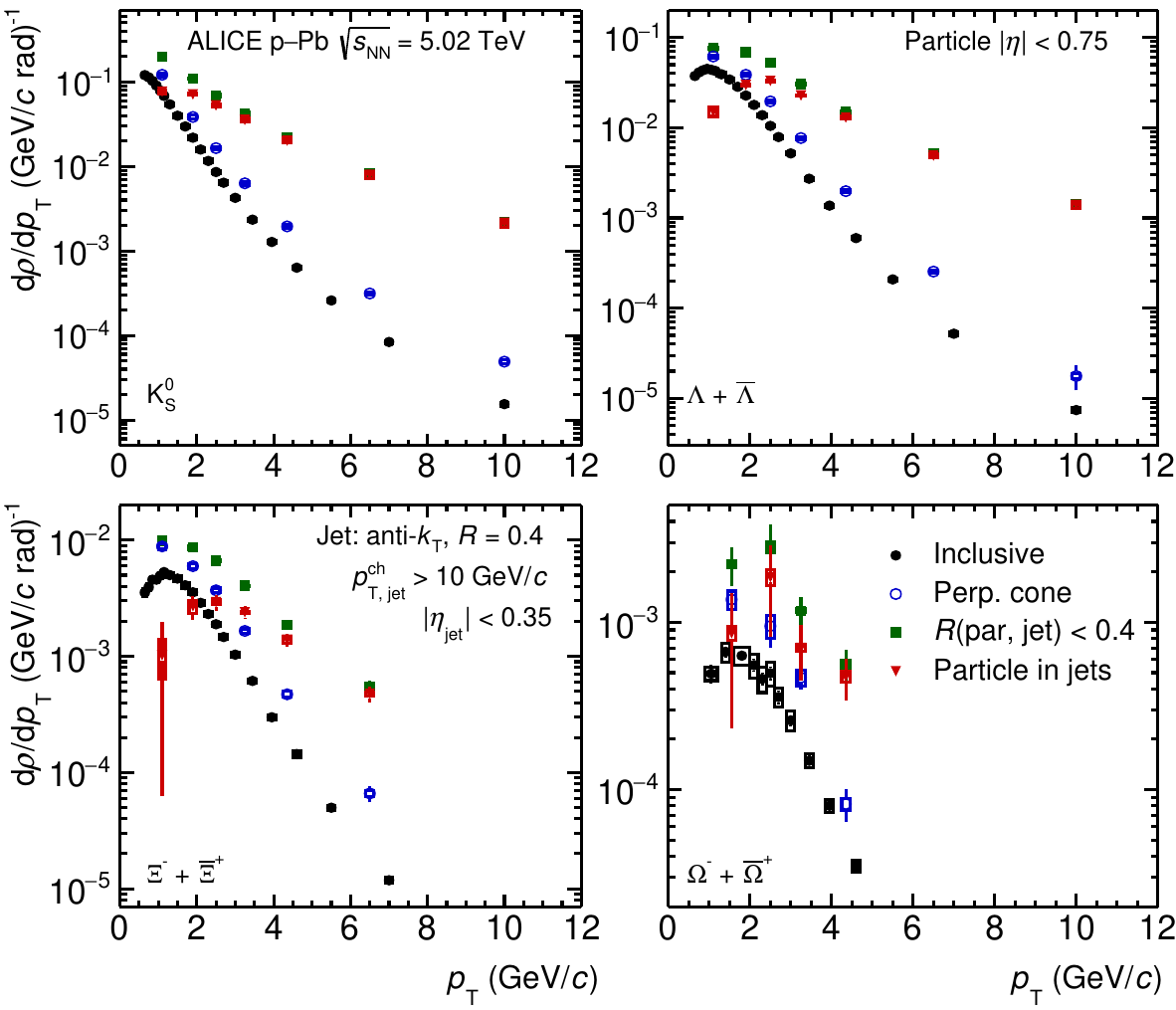}
\end{center}
\caption{$\pT$-differential density, $\dd\rho/\dd\pT$, of $\kzero$ (top left panel), $\lmb$ (top right panel), $\Xi$ (bottom left panel), and $\Omega$ (bottom right panel) in \pPb collisions at \fivenn.
The spectra of JE particles (red triangles), associated with hard scatterings, are compared with that of JC (green squares) and UE (blue open circles) selections.
The results from inclusive measurements (black closed circles) are presented as well.
The statistical uncertainties are represented by the vertical error bars and the systematic uncertainties by the boxes.}
\label{fig:pPbSpect}
\end{figure}

\subsection{Production and ratios of JE particles in \pPb collisions at ${\bf \sqrt{s_{\bf\rm NN}} = 5.02}$~\TeV}

The proton to $\pi^{\pm}$ ratios and strange baryon-to-meson yield ratios measured at high multiplicity in small collision systems (pp and \pPb)~\cite{ALICE:2019hno,ALICE:2013wgn,Khachatryan:2016yru,ALICE:2016dei,ALICE:2020jsh,ALICE:2017jyt,Acharya:2018orn} exhibit an enhancement at intermediate $\pT\sim 3$~\GeVc with respect to the low-multiplicity events, qualitatively reminiscent of that measured in \PbPb collisions~\cite{ALICE:2012ovd,ALICE:2013mez,ALICE:2013cdo,ALICE:2013xmt}.
In the latter, the enhancement is considered as the fingerprint of hydrodynamic evolution of the colour-deconfined matter state, the quark--gluon plasma, created under extreme conditions of high temperature and energy density.
To further constrain the particle production mechanisms in small collision systems, the study of strange particle production within charged-particle jets is extended to \pPb collisions at \fivenn in both MB events and in events selected in various multiplicity intervals.

Figure~\ref{fig:pPbSpect} shows the $\pT$-differential densities of $\kzero$, $\lmb$, $\Xi$, and $\Omega$ in MB \pPb collisions at \fivenn.
For each case, the density distribution of JE particles is compared with that from JC and UE selections, and that of inclusive particles.
In general, the particle densities measured in \pPb collisions have the same order of magnitude as the corresponding ones in pp collisions shown in Fig.~\ref{fig:ppSpect}.
Similar to the case in pp collisions, the density of JC-selected particles is dominated by those associated with hard scattering at high $\pT$ ($\pT > 3$~\GeVc ).
As in pp, the $\pT$-dependent density of JE particles is considerably less steep than the inclusive ones.
The UE component given by the PC selection is mainly located in the low-$\pT$ region but the contribution is larger than in pp collisions.
The UE fractions are $61.3\%$ ($47.0\%$), $80.4\%$ ($65.5\%$), $90.0\%$ ($73.4\%$) and $61.4\%$ ($62.0\%$) for $\kzero$, $\lmb$, and $\Xi$ in the interval of $0.9 < \pT < 1.6$~\GeVc and for $\Omega$ in the interval of $0.9 < \pT < 2.2$~\GeVc in \pPb (pp) collisions, respectively.
This follows the expectation that the multiplicity of particles within the UE is correlated with the number of nucleon$-$nucleon interactions and the energy density of the system.
The $\pT$-dependent densities of $\kzero$ mesons and $\lmb$ and $\Xi$ baryons with different selections are also measured for various event multiplicity classes, presented in Figs.~\ref{fig:cSpect},~\ref{fig:sSpect}, and~\ref{fig:pSpect} in Appendix~\ref{app:pPbCent}.
The magnitude of the density increases with the event multiplicity, and the results in the low multiplicity class become almost identical to those in pp collisions.

\begin{figure}[!t]
\begin{center}
\includegraphics[width=1.\textwidth]{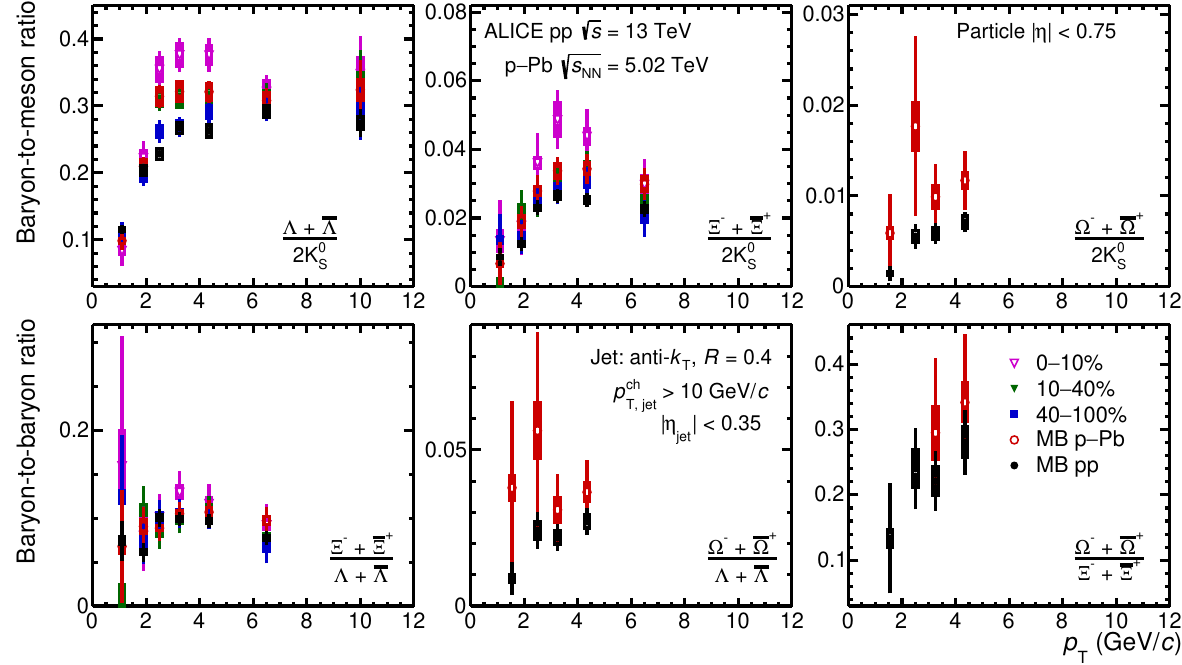}
\end{center}
\caption{$\pT$-dependent strange baryon-to-meson (top) and baryon-to-baryon (bottom) ratios for particles produced within jets in \pPb collisions at \fivenn.
For each case, the results for different event multiplicity classes are compared with that in pp collisions at \thirteen.
The statistical uncertainties are represented by the vertical error bars and the systematic uncertainties by the boxes.
See the text for details.}
\label{fig:pPbRatio}
\end{figure}

The ratios of JE particles measured in \pPb collisions at \fivenn are shown in Fig.~\ref{fig:pPbRatio}.
The results of $\lmb$/$\kzero$, $\Xi$/$\kzero$, and $\Xi$/$\lmb$ ratios are presented in three event multiplicity classes, from high (\cent{0}{10}), intermediate (\cent{10}{40}) to low (\cent{40}{100}) multiplicities, and compared with that in MB events.
Since the multiplicity differential analysis is challenging for $\Omega$ baryons due to the low number of candidates the ratios of $\Omega$/$\kzero$, $\Omega$/$\lmb$ and $\Omega$/$\Xi$ are only given for the MB events.
All ratios are also compared with the corresponding measurements in pp collisions at \thirteen.
Similar to what is observed in Ref.~\cite{Acharya:2021oaa}, the $\lmb$/$\kzero$ ratio obtained in \pPb collisions is systematically higher than that in pp collisions for $2 < \pT < 4$~\GeVc: in this $\pT$ interval the ratio in \pPb collisions also increases with the event multiplicity.
The differences are quantified in terms of standard deviations considering statistical and systematic uncertainties.
The differences are $0.8\sigma$ between MB \pPb collisions and pp collisions and $1.1\sigma$ between high- (\cent{0}{10}) and low-multiplicity (\cent{40}{100}) \pPb collisions.
A similar behavior is observed in ratios between other particle species.
However, due to substantial uncertainties, variations between collision systems or among different event multiplicity classes for \pPb collisions are much less significant than for the $\lmb$/$\kzero$ ratio.
The observed deviations remain to be studied with better statistical precision.

The comparisons of the ratios for JE particles to those for inclusive and UE particles for different event multiplicity classes are shown in Figs.~\ref{fig:cRatio},~\ref{fig:sRatio}, and~\ref{fig:pRatio} in Appendix~\ref{app:pPbCent} for $\lmb$/$\kzero$, $\Xi$/$\kzero$, and $\Xi$/$\lmb$, respectively.
Similar to what is observed in pp collisions, in each event multiplicity class, the $\pT$ dependence of the ratio for UE particles is consistent with that of inclusive particles within uncertainties.
Furthermore, the enhancement of particle ratios at intermediate $\pT$ is not observed in jets, suggesting that the enhancement of particle ratios observed at high multiplicity in small systems is only present in the UE and not a feature of jet fragmentation.

\section{Summary}%
\label{sec:Summary}

The production of $\kzero$ mesons and $\lmb$, $\Xi$, and $\Omega$ baryons is measured separately for particles associated with hard scatterings and the underlying event for the first time at the LHC in pp collisions at \thirteen and \pPb collisions at \fivenn.
The results in pp collisions are compared with \Pyeight CR-BLC simulations.
\Pyeight simulations reproduce fairly well the $\lmb$/$\kzero$ ratio in data.
However, large discrepancies between data and simulations are observed in the ratios including multi-strange baryons.
In \pPb collisions, the strange baryon-to-meson and baryon-to-baryon yield ratios associated with jets for different event multiplicity classes have a similar trend.
Better statistical precision is required to clarify their multiplicity and collision system dependence.
The enhancement in the ratio at intermediate $\pT$ found in the inclusive particle measurements in high-multiplicity \pPb\ and \PbPb\ collisions is not present for particles associated with hard scatterings selected by jets reconstructed from charged particles for $\pTjch>10$~\GeVc.
Moreover, as the enhancement has been linked to the interplay of radial flow and parton recombination at intermediate $\pT$, its absence within the jet cone demonstrates that these effects are indeed limited to the soft particle production processes.


\newenvironment{acknowledgement}{\relax}{\relax}
\begin{acknowledgement}
\section*{Acknowledgements}

The ALICE Collaboration would like to thank all its engineers and technicians for their invaluable contributions to the construction of the experiment and the CERN accelerator teams for the outstanding performance of the LHC complex.
The ALICE Collaboration gratefully acknowledges the resources and support provided by all Grid centres and the Worldwide LHC Computing Grid (WLCG) collaboration.
The ALICE Collaboration acknowledges the following funding agencies for their support in building and running the ALICE detector:
A. I. Alikhanyan National Science Laboratory (Yerevan Physics Institute) Foundation (ANSL), State Committee of Science and World Federation of Scientists (WFS), Armenia;
Austrian Academy of Sciences, Austrian Science Fund (FWF): [M 2467-N36] and Nationalstiftung f\"{u}r Forschung, Technologie und Entwicklung, Austria;
Ministry of Communications and High Technologies, National Nuclear Research Center, Azerbaijan;
Conselho Nacional de Desenvolvimento Cient\'{\i}fico e Tecnol\'{o}gico (CNPq), Financiadora de Estudos e Projetos (Finep), Funda\c{c}\~{a}o de Amparo \`{a} Pesquisa do Estado de S\~{a}o Paulo (FAPESP) and Universidade Federal do Rio Grande do Sul (UFRGS), Brazil;
Bulgarian Ministry of Education and Science, within the National Roadmap for Research Infrastructures 2020-2027 (object CERN), Bulgaria;
Ministry of Education of China (MOEC) , Ministry of Science \& Technology of China (MSTC) and National Natural Science Foundation of China (NSFC), China;
Ministry of Science and Education and Croatian Science Foundation, Croatia;
Centro de Aplicaciones Tecnol\'{o}gicas y Desarrollo Nuclear (CEADEN), Cubaenerg\'{\i}a, Cuba;
Ministry of Education, Youth and Sports of the Czech Republic, Czech Republic;
The Danish Council for Independent Research | Natural Sciences, the VILLUM FONDEN and Danish National Research Foundation (DNRF), Denmark;
Helsinki Institute of Physics (HIP), Finland;
Commissariat \`{a} l'Energie Atomique (CEA) and Institut National de Physique Nucl\'{e}aire et de Physique des Particules (IN2P3) and Centre National de la Recherche Scientifique (CNRS), France;
Bundesministerium f\"{u}r Bildung und Forschung (BMBF) and GSI Helmholtzzentrum f\"{u}r Schwerionenforschung GmbH, Germany;
General Secretariat for Research and Technology, Ministry of Education, Research and Religions, Greece;
National Research, Development and Innovation Office, Hungary;
Department of Atomic Energy Government of India (DAE), Department of Science and Technology, Government of India (DST), University Grants Commission, Government of India (UGC) and Council of Scientific and Industrial Research (CSIR), India;
National Research and Innovation Agency - BRIN, Indonesia;
Istituto Nazionale di Fisica Nucleare (INFN), Italy;
Japanese Ministry of Education, Culture, Sports, Science and Technology (MEXT) and Japan Society for the Promotion of Science (JSPS) KAKENHI, Japan;
Consejo Nacional de Ciencia (CONACYT) y Tecnolog\'{i}a, through Fondo de Cooperaci\'{o}n Internacional en Ciencia y Tecnolog\'{i}a (FONCICYT) and Direcci\'{o}n General de Asuntos del Personal Academico (DGAPA), Mexico;
Nederlandse Organisatie voor Wetenschappelijk Onderzoek (NWO), Netherlands;
The Research Council of Norway, Norway;
Commission on Science and Technology for Sustainable Development in the South (COMSATS), Pakistan;
Pontificia Universidad Cat\'{o}lica del Per\'{u}, Peru;
Ministry of Education and Science, National Science Centre and WUT ID-UB, Poland;
Korea Institute of Science and Technology Information and National Research Foundation of Korea (NRF), Republic of Korea;
Ministry of Education and Scientific Research, Institute of Atomic Physics, Ministry of Research and Innovation and Institute of Atomic Physics and University Politehnica of Bucharest, Romania;
Ministry of Education, Science, Research and Sport of the Slovak Republic, Slovakia;
National Research Foundation of South Africa, South Africa;
Swedish Research Council (VR) and Knut \& Alice Wallenberg Foundation (KAW), Sweden;
European Organization for Nuclear Research, Switzerland;
Suranaree University of Technology (SUT), National Science and Technology Development Agency (NSTDA), Thailand Science Research and Innovation (TSRI) and National Science, Research and Innovation Fund (NSRF), Thailand;
Turkish Energy, Nuclear and Mineral Research Agency (TENMAK), Turkey;
National Academy of  Sciences of Ukraine, Ukraine;
Science and Technology Facilities Council (STFC), United Kingdom;
National Science Foundation of the United States of America (NSF) and United States Department of Energy, Office of Nuclear Physics (DOE NP), United States of America.
In addition, individual groups or members have received support from:
Marie Sk\l{}odowska Curie, European Research Council, Strong 2020 - Horizon 2020 (grant nos. 950692, 824093, 896850), European Union;
Academy of Finland (Center of Excellence in Quark Matter) (grant nos. 346327, 346328), Finland;
Programa de Apoyos para la Superaci\'{o}n del Personal Acad\'{e}mico, UNAM, Mexico.

\end{acknowledgement}

\bibliographystyle{etc/utphys}
\bibliography{AliStrangeJets}

\newpage
\appendix

\section{Particle candidate selection criteria}%
\label{app:cuts}

\begin{table}[!ht]
\begin{center}
\caption{\kzero(\lmb and \almb) candidate selection criteria of topological variables, daughter tracks and \Vzero candidates.
The DCA stands for the ``distance of closest approach'', PV represents the ``primary collision vertex'' and CPA is the ``cosine pointing angle between the momentum vector of the reconstructed \Vzero and the displacement vector between the decay and primary vertices''.}
\label{tab:V0Cut}
\begin{tabularx}{\textwidth}{@{} lCC @{}}
\toprule
\textbf{Topological variable} & \textbf{\pp} & \textbf{\pPb} \\
\midrule
$\Vzero$ transverse decay radius      & $> 0.5$~cm   & $> 0.5$~cm \\
DCA of $\Vzero$ daughter track to PV & $> 0.06$~cm  & $> 0.06$~cm \\
DCA between $\Vzero$ daughter tracks  & $< 1\sigma$ & $< 1\sigma$ \\
CPA of $\Vzero$ & $> 0.97$ ($0.995$) & $> 0.97$ ($0.995$) \\
\midrule
\textbf{Track selection} \\
\midrule
Daughter track pseudorapidity interval &$|\eta| < 0.8$ & $\abs{\eta} < 0.8$      \\
Daughter track $N_{\rm crossed~rows}$                   & $\geq 70$  & $\geq$ 70 \\
Daughter track $N_{\rm crossed~rows}/N_{\rm findable}$  & $\geq 0.8$ & $\geq$ 0.8 \\
TPC $\dEdx$ & $< 5\sigma$ & $< 5\sigma$ \\
\midrule
\textbf{Candidate selection} \\
\midrule
Pseudorapidity interval & $|\eta| < 0.75$ & $|\eta| < 0.75$ \\
Proper decay length & $< 20$ (30)~cm & $< 20$ (30)~cm \\
Competing mass & $> 0.005$ (0.010)~\GeVmass & $> 0.005$ (0.010)~\GeVmass \\
\bottomrule
\end{tabularx}
\end{center}
\end{table}

\begin{table}[!ht]
\begin{center}
\caption{\Xis(\Oms) candidate selection criteria of topological variables, daughter tracks and cascade candidates.}
\label{tab:CascadeCut}
\begin{tabularx}{\textwidth}{@{} lCC @{}}
\toprule
\textbf{Topological variable} & \textbf{\pp} & \textbf{\pPb} \\
\midrule
Cascade transverse decay radius & $> 0.8(0.6)$~cm & $> 0.6$~cm \\
\Vzero transverse decay radius & $> 1.4$~cm     & $> 1.2$~cm \\
DCA (bachelor to PV)           & $> 0.05$~cm    & $> 0.04$~cm \\
DCA (\Vzero to PV)             & $> 0.07$~cm    & $> 0.06$~cm \\
DCA (positive / negative track to PV) & $> 0.04(0.03)$~cm & $> 0.03$~cm  \\
DCA between \Vzero daughter tracks & $< 1.6\sigma$     & $< 1.5\sigma$ \\
DCA (bachelor to \Vzero) & $< 1.6(1.0)$~cm & $< 1.3$~cm \\
CPA of Cascade          & $> 0.97$       & $> 0.97$  \\
CPA of \Vzero           & $> 0.97$       & $> 0.97$  \\
\Vzero invariant mass window & $\pm 0.006$~\GeVmass & $\pm 0.008$~\GeVmass \\
\midrule
\textbf{Track selection} \\
\midrule
Daughter track pseudorapidity interval & $|\eta| < 0.8$ & $|\eta| < 0.8$ \\
Daughter track $N_{\rm crossed~rows}$  & $\geq 70$      & $\geq$ 70 \\
Daughter track $N_{\rm crossed~rows}/N_{\rm findable}$ &$\geq 0.8$ &$\geq$ 0.8 \\
TPC $\dEdx$ & $< 5\sigma$ & $< 4\sigma$ \\
\midrule
\textbf{Candidate selection} \\
\midrule
Pseudorapidity interval & $|\eta| < 0.75$ &$|\eta| < 0.75$ \\
Proper decay length & -- & $< 3 \times$ mean decay length \\
Competing mass          & $8$~\MeVmass & $8$~\MeVmass \\
\bottomrule
\end{tabularx}
\end{center}
\end{table}

\clearpage
\section{Comparison of $\Xi$/$\lmb$ and $\Omega$/$\lmb$ ratios to colour-rope predictions in pp collisions at \thirteen}%
\label{app:ppRope}

\begin{figure}[!h]
\begin{center}
\includegraphics[width=.48\textwidth]{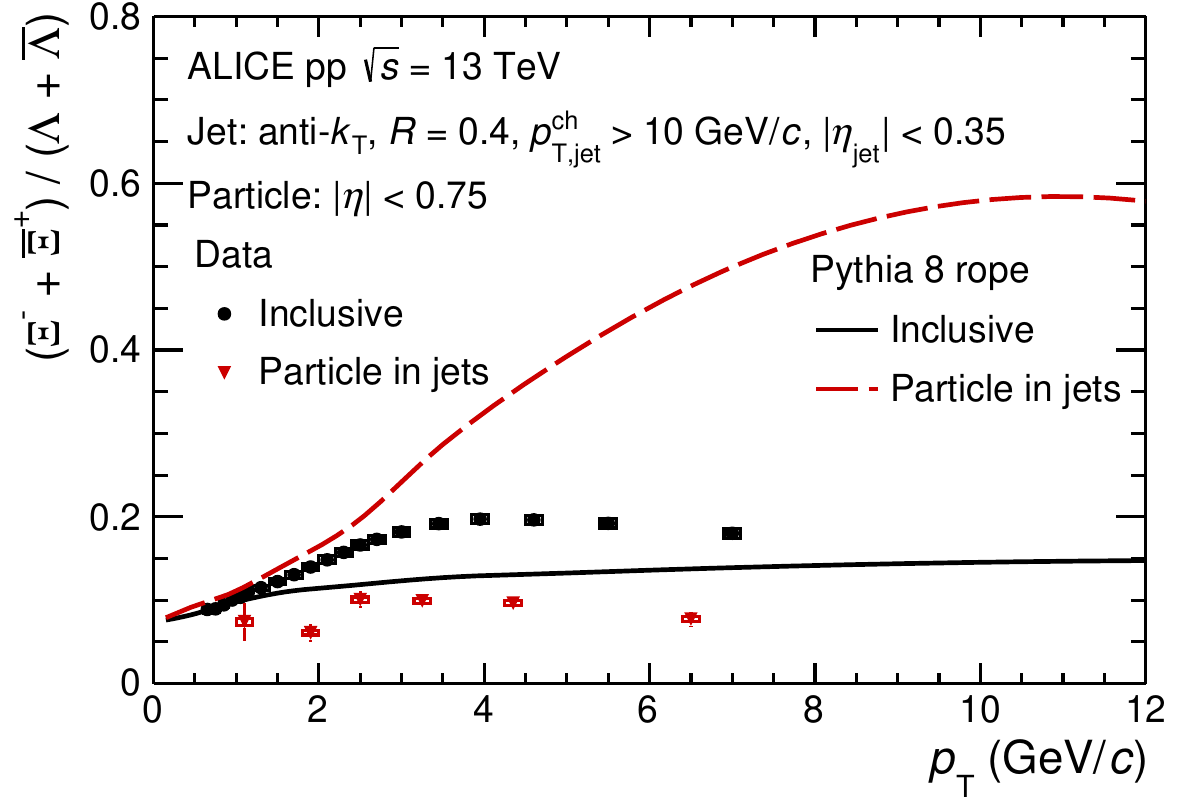}
\includegraphics[width=.48\textwidth]{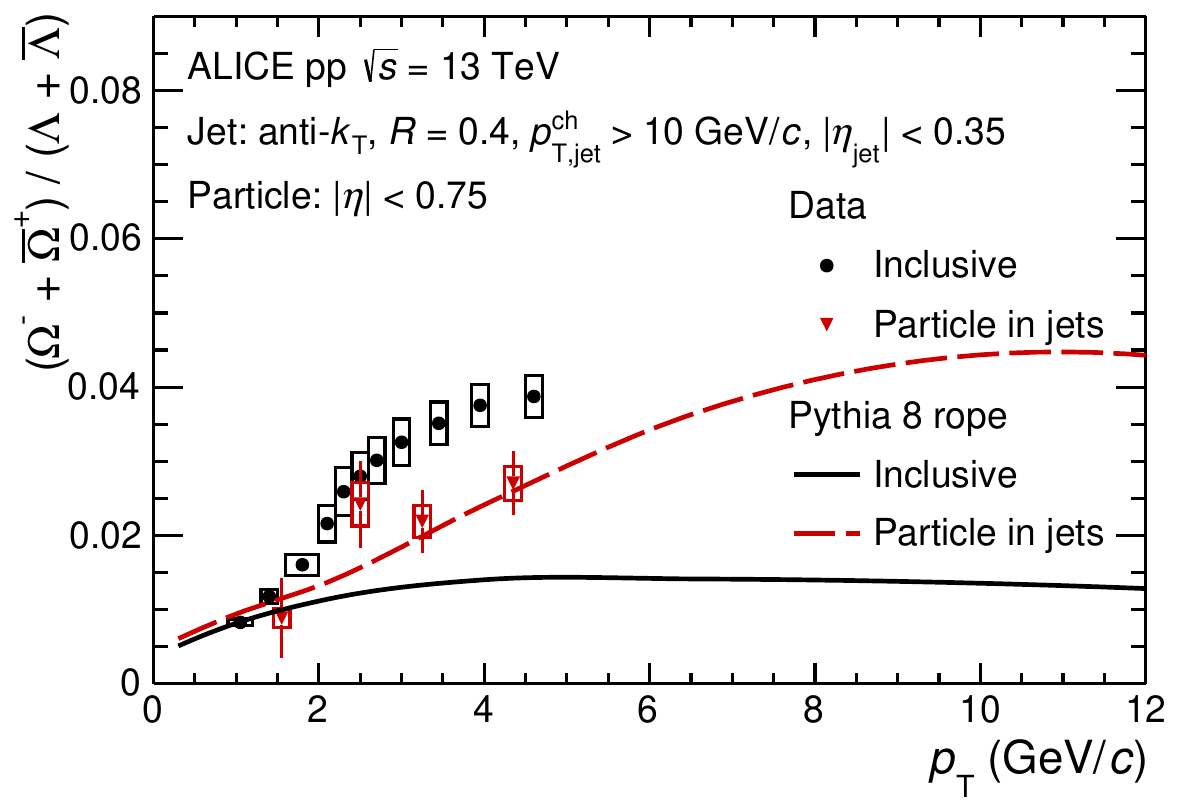}
\end{center}
\caption{$\pT$-dependent $\Xi$/$\lmb$ (left panel) and $\Omega$/$\lmb$ (right panel) ratios in pp collisions at \thirteen.
For each case, the results of JE particles (red inverted-triangle) are compared with that of inclusive (black closed circles).
The corresponding colour-rope predictions~\cite{Bierlich:2014xba} implemented in the \Pyeight event generator~\cite{Bierlich:2022pfr} are compared to data as well.}
\label{fig:Rope}
\end{figure}

\clearpage
\section{$\pT$-differential particle density and ratios for event multiplicity classes in \pPb collisions at \fivenn}%
\label{app:pPbCent}

\begin{figure}[!h]
\begin{center}
\includegraphics[width=\textwidth]{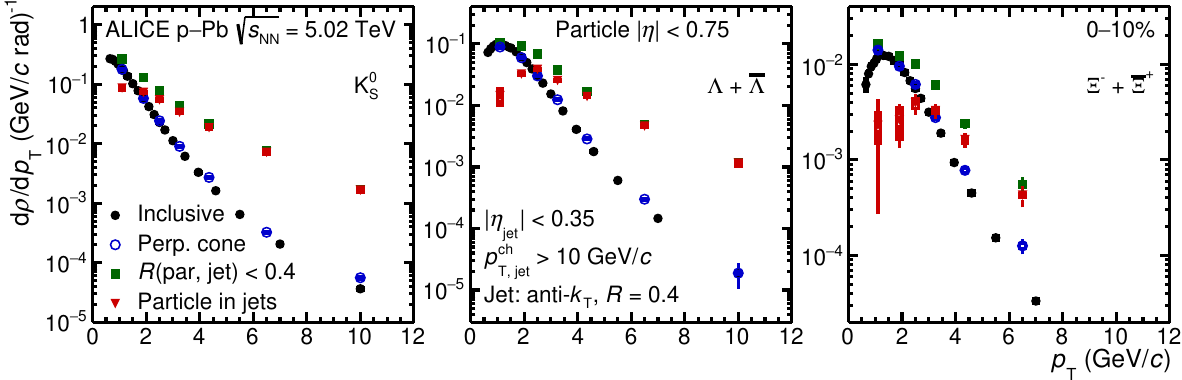}
\end{center}
\caption{$\pT$-differential density, $\dd\rho/\dd\pT$, of $\kzero$ (left panel), $\lmb$ (middle panel) and $\Xi$ (right panel) for the \cent{0}{10} event multiplicity class in \pPb collisions at \fivenn.
The spectra of JE particles (red inverted-triangle), associated with hard scatterings, are compared with that of JC (green squares) and UE (blue open circles) selections.
The results from inclusive measurements (black closed circles) are presented as well.
The statistical uncertainties are represented by the vertical error bars and the systematic uncertainties by the boxes.}
\label{fig:cSpect}
\end{figure}

\begin{figure}[!h]
\begin{center}
\includegraphics[width=\textwidth]{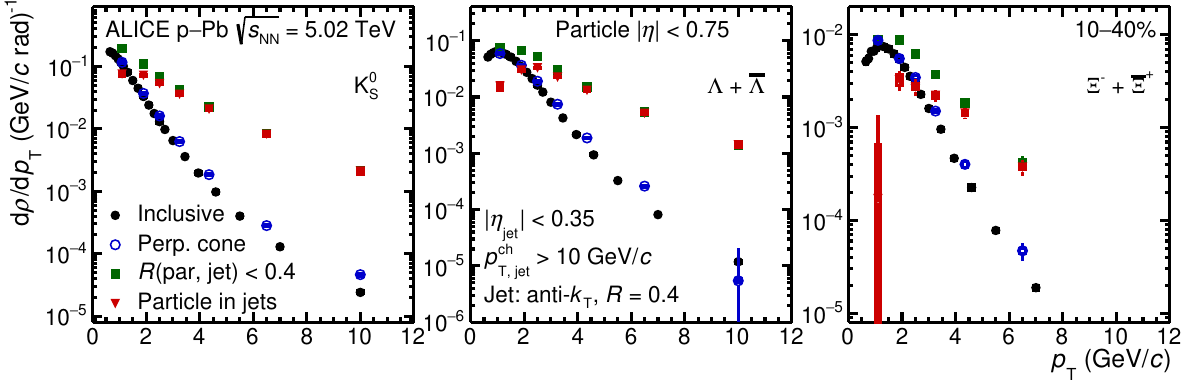}
\end{center}
\caption{$\pT$-differential density, $\dd\rho/\dd\pT$, of $\kzero$ (left panel), $\lmb$ (middle panel) and $\Xi$ (right panel) for the \cent{10}{40} event multiplicity class in \pPb collisions at \fivenn.
The spectra of JE particles (red inverted-triangle), associated with hard scatterings, are compared with that of JC (green squares) and UE (blue open circles) selections.
The results from inclusive measurements (black closed circles) are presented as well.
The statistical uncertainties are represented by the vertical error bars and the systematic uncertainties by the boxes.}
\label{fig:sSpect}
\end{figure}

\begin{figure}[!h]
\begin{center}
\includegraphics[width=\textwidth]{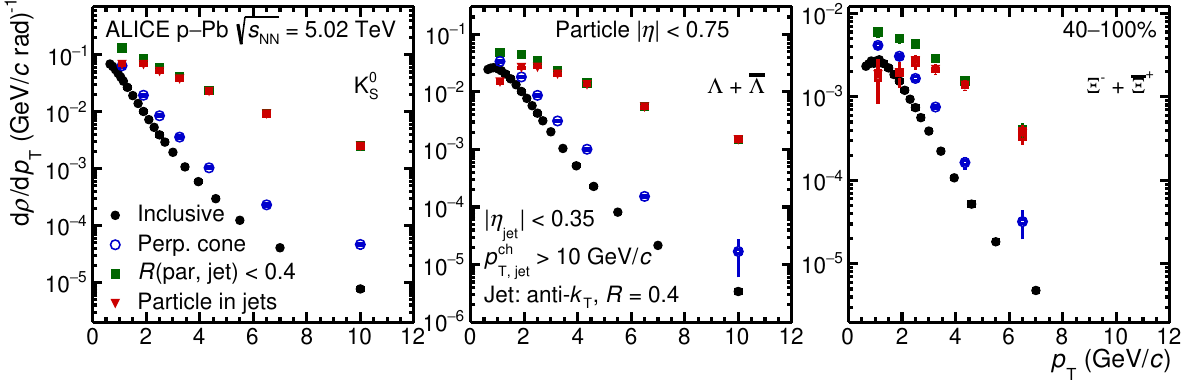}
\end{center}
\caption{$\pT$-differential density, $\dd\rho/\dd\pT$, of $\kzero$ (left panel), $\lmb$ (middle panel) and $\Xi$ (right panel) for the \cent{40}{100} event multiplicity class in \pPb collisions at \fivenn.
The spectra of JE particles (red inverted-triangle), associated with hard scatterings, are compared with that of JC (green squares) and UE (blue open circles) selections.
The results from inclusive measurements (black closed circles) are presented as well.
The statistical uncertainties are represented by the vertical error bars and the systematic uncertainties by the boxes.}
\label{fig:pSpect}
\end{figure}

\begin{figure}[!h]
\begin{center}
\includegraphics[width=\textwidth]{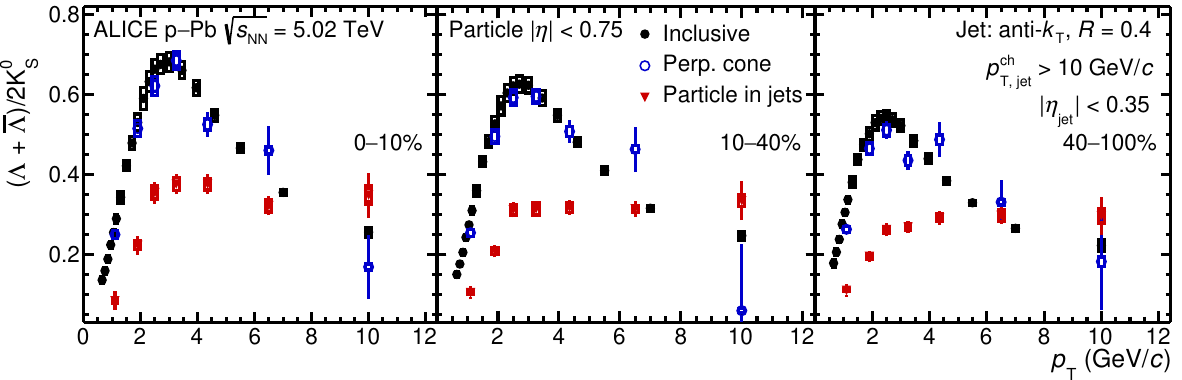}
\end{center}
\caption{$\pT$-dependent $\lmb$/$\kzero$ ratio for \cent{0}{10} (left panel), \cent{10}{40} (middle panel) and \cent{40}{100} (right panel) event multiplicity classes.
For each case, the results of JE particles (red inverted-triangle) are compared with that of inclusive (black closed circles) and UE (blue open circles) particles.
The statistical uncertainties are represented by the vertical error bars and the systematic uncertainties by the boxes.}
\label{fig:cRatio}
\end{figure}

\begin{figure}[!h]
\begin{center}
\includegraphics[width=\textwidth]{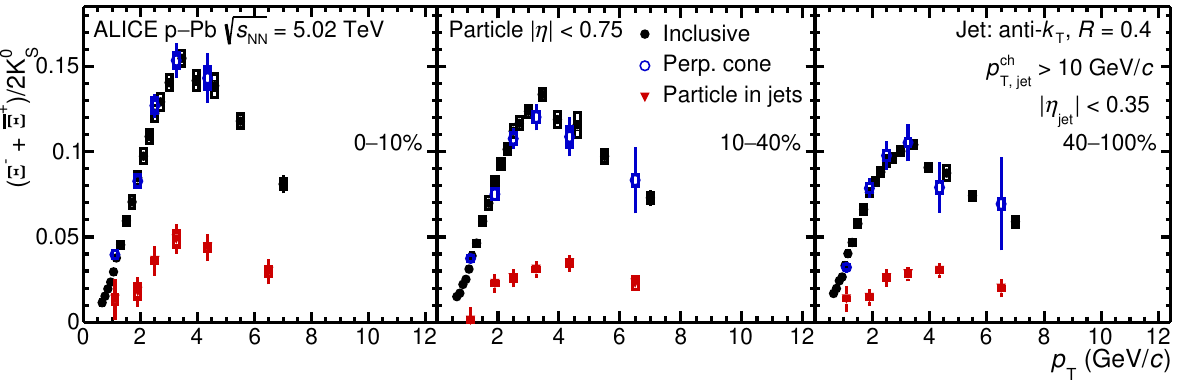}
\end{center}
\caption{$\pT$-dependent $\Xi$/$\kzero$ ratio for \cent{0}{10} (left panel), \cent{10}{40} (middle panel) and \cent{40}{100} (right panel) event multiplicity classes.
For each case, the results of JE particles (red inverted-triangle) are compared with that of inclusive (black closed circles) and UE (blue open circles) particles.
The statistical uncertainties are represented by the vertical error bars and the systematic uncertainties by the boxes.}
\label{fig:sRatio}
\end{figure}

\begin{figure}[!h]
\begin{center}
\includegraphics[width=\textwidth]{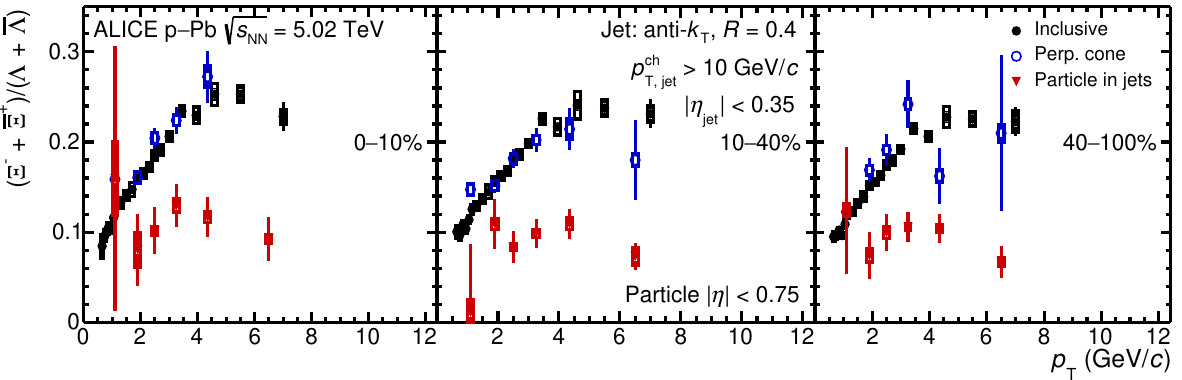}
\end{center}
\caption{$\pT$-dependent $\Xi$/$\lmb$ ratio for \cent{0}{10} (left panel), \cent{10}{40} (middle panel) and \cent{40}{100} (right panel) event multiplicity classes.
For each case, the results of JE particles (red inverted-triangle) are compared with that of inclusive (black closed circles) and UE (blue open circles) particles.
The statistical uncertainties are represented by the vertical error bars and the systematic uncertainties by the boxes.}
\label{fig:pRatio}
\end{figure}

\clearpage

\section{The ALICE Collaboration}
\label{app:collab}
\begin{flushleft} 
\small

S.~Acharya\,\orcidlink{0000-0002-9213-5329}\,$^{\rm 125}$, 
D.~Adamov\'{a}\,\orcidlink{0000-0002-0504-7428}\,$^{\rm 86}$, 
A.~Adler$^{\rm 69}$, 
G.~Aglieri Rinella\,\orcidlink{0000-0002-9611-3696}\,$^{\rm 32}$, 
M.~Agnello\,\orcidlink{0000-0002-0760-5075}\,$^{\rm 29}$, 
N.~Agrawal\,\orcidlink{0000-0003-0348-9836}\,$^{\rm 50}$, 
Z.~Ahammed\,\orcidlink{0000-0001-5241-7412}\,$^{\rm 132}$, 
S.~Ahmad\,\orcidlink{0000-0003-0497-5705}\,$^{\rm 15}$, 
S.U.~Ahn\,\orcidlink{0000-0001-8847-489X}\,$^{\rm 70}$, 
I.~Ahuja\,\orcidlink{0000-0002-4417-1392}\,$^{\rm 37}$
% ...

\section*{Affiliation Notes}

$^{\rm I}$ Deceased\\
$^{\rm II}$ Also at: Max-Planck-Institut f\"{u}r Physik, Munich, Germany\\
$^{\rm III}$ Also at: Italian National Agency for New Technologies, Energy and Sustainable Economic Development (ENEA), Bologna, Italy\\
$^{\rm IV}$ Also at: Dipartimento DET del Politecnico di Torino, Turin, Italy\\
$^{\rm V}$ Also at: Department of Applied Physics, Aligarh Muslim University, Aligarh, India\\
$^{\rm VI}$ Also at: Institute of Theoretical Physics, University of Wroclaw, Poland\\
$^{\rm VII}$ Also at: An institution covered by a cooperation agreement with CERN\\

\section*{Collaboration Institutes}

$^{1}$ A.I. Alikhanyan National Science Laboratory (Yerevan Physics Institute) Foundation, Yerevan, Armenia\\
$^{2}$ AGH University of Science and Technology, Cracow, Poland\\
$^{3}$ Bogolyubov Institute for Theoretical Physics, National Academy of Sciences of Ukraine, Kiev, Ukraine\\
$^{4}$ Bose Institute, Department of Physics  and Centre for Astroparticle Physics and Space Science (CAPSS), Kolkata, India\\
$^{5}$ California Polytechnic State University, San Luis Obispo, California, United States\\
$^{6}$ Central China Normal University, Wuhan, China\\
$^{7}$ Centro de Aplicaciones Tecnol\'{o}gicas y Desarrollo Nuclear (CEADEN), Havana, Cuba\\
$^{8}$ Centro de Investigaci\'{o}n y de Estudios Avanzados (CINVESTAV), Mexico City and M\'{e}rida, Mexico\\
$^{9}$ Chicago State University, Chicago, Illinois, United States\\
$^{10}$ China Institute of Atomic Energy, Beijing, China\\
$^{11}$ Chungbuk National University, Cheongju, Republic of Korea\\
$^{12}$ Comenius University Bratislava, Faculty of Mathematics, Physics and Informatics, Bratislava, Slovak Republic\\
$^{13}$ COMSATS University Islamabad, Islamabad, Pakistan\\
$^{14}$ Creighton University, Omaha, Nebraska, United States\\
$^{15}$ Department of Physics, Aligarh Muslim University, Aligarh, India\\
$^{16}$ Department of Physics, Pusan National University, Pusan, Republic of Korea\\
$^{17}$ Department of Physics, Sejong University, Seoul, Republic of Korea\\
$^{18}$ Department of Physics, University of California, Berkeley, California, United States\\
$^{19}$ Department of Physics, University of Oslo, Oslo, Norway\\
$^{20}$ Department of Physics and Technology, University of Bergen, Bergen, Norway\\
$^{21}$ Dipartimento di Fisica, Universit\`{a} di Pavia, Pavia, Italy\\
$^{22}$ Dipartimento di Fisica dell'Universit\`{a} and Sezione INFN, Cagliari, Italy\\
$^{23}$ Dipartimento di Fisica dell'Universit\`{a} and Sezione INFN, Trieste, Italy\\
$^{24}$ Dipartimento di Fisica dell'Universit\`{a} and Sezione INFN, Turin, Italy\\
$^{25}$ Dipartimento di Fisica e Astronomia dell'Universit\`{a} and Sezione INFN, Bologna, Italy\\
$^{26}$ Dipartimento di Fisica e Astronomia dell'Universit\`{a} and Sezione INFN, Catania, Italy\\
$^{27}$ Dipartimento di Fisica e Astronomia dell'Universit\`{a} and Sezione INFN, Padova, Italy\\
$^{28}$ Dipartimento di Fisica `E.R.~Caianiello' dell'Universit\`{a} and Gruppo Collegato INFN, Salerno, Italy\\
$^{29}$ Dipartimento DISAT del Politecnico and Sezione INFN, Turin, Italy\\
$^{30}$ Dipartimento di Scienze MIFT, Universit\`{a} di Messina, Messina, Italy\\
$^{31}$ Dipartimento Interateneo di Fisica `M.~Merlin' and Sezione INFN, Bari, Italy\\
$^{32}$ European Organization for Nuclear Research (CERN), Geneva, Switzerland\\
$^{33}$ Faculty of Electrical Engineering, Mechanical Engineering and Naval Architecture, University of Split, Split, Croatia\\
$^{34}$ Faculty of Engineering and Science, Western Norway University of Applied Sciences, Bergen, Norway\\
$^{35}$ Faculty of Nuclear Sciences and Physical Engineering, Czech Technical University in Prague, Prague, Czech Republic\\
$^{36}$ Faculty of Physics, Sofia University, Sofia, Bulgaria\\
$^{37}$ Faculty of Science, P.J.~\v{S}af\'{a}rik University, Ko\v{s}ice, Slovak Republic\\
$^{38}$ Frankfurt Institute for Advanced Studies, Johann Wolfgang Goethe-Universit\"{a}t Frankfurt, Frankfurt, Germany\\
$^{39}$ Fudan University, Shanghai, China\\
$^{40}$ Gangneung-Wonju National University, Gangneung, Republic of Korea\\
$^{41}$ Gauhati University, Department of Physics, Guwahati, India\\
$^{42}$ Helmholtz-Institut f\"{u}r Strahlen- und Kernphysik, Rheinische Friedrich-Wilhelms-Universit\"{a}t Bonn, Bonn, Germany\\
$^{43}$ Helsinki Institute of Physics (HIP), Helsinki, Finland\\
$^{44}$ High Energy Physics Group,  Universidad Aut\'{o}noma de Puebla, Puebla, Mexico\\
$^{45}$ Horia Hulubei National Institute of Physics and Nuclear Engineering, Bucharest, Romania\\
$^{46}$ Indian Institute of Technology Bombay (IIT), Mumbai, India\\
$^{47}$ Indian Institute of Technology Indore, Indore, India\\
$^{48}$ INFN, Laboratori Nazionali di Frascati, Frascati, Italy\\
$^{49}$ INFN, Sezione di Bari, Bari, Italy\\
$^{50}$ INFN, Sezione di Bologna, Bologna, Italy\\
$^{51}$ INFN, Sezione di Cagliari, Cagliari, Italy\\
$^{52}$ INFN, Sezione di Catania, Catania, Italy\\
$^{53}$ INFN, Sezione di Padova, Padova, Italy\\
$^{54}$ INFN, Sezione di Pavia, Pavia, Italy\\
$^{55}$ INFN, Sezione di Torino, Turin, Italy\\
$^{56}$ INFN, Sezione di Trieste, Trieste, Italy\\
$^{57}$ Inha University, Incheon, Republic of Korea\\
$^{58}$ Institute for Gravitational and Subatomic Physics (GRASP), Utrecht University/Nikhef, Utrecht, Netherlands\\
$^{59}$ Institute of Experimental Physics, Slovak Academy of Sciences, Ko\v{s}ice, Slovak Republic\\
$^{60}$ Institute of Physics, Homi Bhabha National Institute, Bhubaneswar, India\\
$^{61}$ Institute of Physics of the Czech Academy of Sciences, Prague, Czech Republic\\
$^{62}$ Institute of Space Science (ISS), Bucharest, Romania\\
$^{63}$ Institut f\"{u}r Kernphysik, Johann Wolfgang Goethe-Universit\"{a}t Frankfurt, Frankfurt, Germany\\
$^{64}$ Instituto de Ciencias Nucleares, Universidad Nacional Aut\'{o}noma de M\'{e}xico, Mexico City, Mexico\\
$^{65}$ Instituto de F\'{i}sica, Universidade Federal do Rio Grande do Sul (UFRGS), Porto Alegre, Brazil\\
$^{66}$ Instituto de F\'{\i}sica, Universidad Nacional Aut\'{o}noma de M\'{e}xico, Mexico City, Mexico\\
$^{67}$ iThemba LABS, National Research Foundation, Somerset West, South Africa\\
$^{68}$ Jeonbuk National University, Jeonju, Republic of Korea\\
$^{69}$ Johann-Wolfgang-Goethe Universit\"{a}t Frankfurt Institut f\"{u}r Informatik, Fachbereich Informatik und Mathematik, Frankfurt, Germany\\
$^{70}$ Korea Institute of Science and Technology Information, Daejeon, Republic of Korea\\
$^{71}$ KTO Karatay University, Konya, Turkey\\
$^{72}$ Laboratoire de Physique des 2 Infinis, Ir\`{e}ne Joliot-Curie, Orsay, France\\
$^{73}$ Laboratoire de Physique Subatomique et de Cosmologie, Universit\'{e} Grenoble-Alpes, CNRS-IN2P3, Grenoble, France\\
$^{74}$ Lawrence Berkeley National Laboratory, Berkeley, California, United States\\
$^{75}$ Lund University Department of Physics, Division of Particle Physics, Lund, Sweden\\
$^{76}$ Nagasaki Institute of Applied Science, Nagasaki, Japan\\
$^{77}$ Nara Women{'}s University (NWU), Nara, Japan\\
$^{78}$ National and Kapodistrian University of Athens, School of Science, Department of Physics , Athens, Greece\\
$^{79}$ National Centre for Nuclear Research, Warsaw, Poland\\
$^{80}$ National Institute of Science Education and Research, Homi Bhabha National Institute, Jatni, India\\
$^{81}$ National Nuclear Research Center, Baku, Azerbaijan\\
$^{82}$ National Research and Innovation Agency - BRIN, Jakarta, Indonesia\\
$^{83}$ Niels Bohr Institute, University of Copenhagen, Copenhagen, Denmark\\
$^{84}$ Nikhef, National institute for subatomic physics, Amsterdam, Netherlands\\
$^{85}$ Nuclear Physics Group, STFC Daresbury Laboratory, Daresbury, United Kingdom\\
$^{86}$ Nuclear Physics Institute of the Czech Academy of Sciences, Husinec-\v{R}e\v{z}, Czech Republic\\
$^{87}$ Oak Ridge National Laboratory, Oak Ridge, Tennessee, United States\\
$^{88}$ Ohio State University, Columbus, Ohio, United States\\
$^{89}$ Physics department, Faculty of science, University of Zagreb, Zagreb, Croatia\\
$^{90}$ Physics Department, Panjab University, Chandigarh, India\\
$^{91}$ Physics Department, University of Jammu, Jammu, India\\
$^{92}$ Physics Program and International Institute for Sustainability with Knotted Chiral Meta Matter (SKCM2), Hiroshima University, Hiroshima, Japan\\
$^{93}$ Physikalisches Institut, Eberhard-Karls-Universit\"{a}t T\"{u}bingen, T\"{u}bingen, Germany\\
$^{94}$ Physikalisches Institut, Ruprecht-Karls-Universit\"{a}t Heidelberg, Heidelberg, Germany\\
$^{95}$ Physik Department, Technische Universit\"{a}t M\"{u}nchen, Munich, Germany\\
$^{96}$ Politecnico di Bari and Sezione INFN, Bari, Italy\\
$^{97}$ Research Division and ExtreMe Matter Institute EMMI, GSI Helmholtzzentrum f\"ur Schwerionenforschung GmbH, Darmstadt, Germany\\
$^{98}$ Saga University, Saga, Japan\\
$^{99}$ Saha Institute of Nuclear Physics, Homi Bhabha National Institute, Kolkata, India\\
$^{100}$ School of Physics and Astronomy, University of Birmingham, Birmingham, United Kingdom\\
$^{101}$ Secci\'{o}n F\'{\i}sica, Departamento de Ciencias, Pontificia Universidad Cat\'{o}lica del Per\'{u}, Lima, Peru\\
$^{102}$ Stefan Meyer Institut f\"{u}r Subatomare Physik (SMI), Vienna, Austria\\
$^{103}$ SUBATECH, IMT Atlantique, Nantes Universit\'{e}, CNRS-IN2P3, Nantes, France\\
$^{104}$ Sungkyunkwan University, Suwon City, Republic of Korea\\
$^{105}$ Suranaree University of Technology, Nakhon Ratchasima, Thailand\\
$^{106}$ Technical University of Ko\v{s}ice, Ko\v{s}ice, Slovak Republic\\
$^{107}$ The Henryk Niewodniczanski Institute of Nuclear Physics, Polish Academy of Sciences, Cracow, Poland\\
$^{108}$ The University of Texas at Austin, Austin, Texas, United States\\
$^{109}$ Universidad Aut\'{o}noma de Sinaloa, Culiac\'{a}n, Mexico\\
$^{110}$ Universidade de S\~{a}o Paulo (USP), S\~{a}o Paulo, Brazil\\
$^{111}$ Universidade Estadual de Campinas (UNICAMP), Campinas, Brazil\\
$^{112}$ Universidade Federal do ABC, Santo Andre, Brazil\\
$^{113}$ University of Cape Town, Cape Town, South Africa\\
$^{114}$ University of Houston, Houston, Texas, United States\\
$^{115}$ University of Jyv\"{a}skyl\"{a}, Jyv\"{a}skyl\"{a}, Finland\\
$^{116}$ University of Kansas, Lawrence, Kansas, United States\\
$^{117}$ University of Liverpool, Liverpool, United Kingdom\\
$^{118}$ University of Science and Technology of China, Hefei, China\\
$^{119}$ University of South-Eastern Norway, Kongsberg, Norway\\
$^{120}$ University of Tennessee, Knoxville, Tennessee, United States\\
$^{121}$ University of the Witwatersrand, Johannesburg, South Africa\\
$^{122}$ University of Tokyo, Tokyo, Japan\\
$^{123}$ University of Tsukuba, Tsukuba, Japan\\
$^{124}$ University Politehnica of Bucharest, Bucharest, Romania\\
$^{125}$ Universit\'{e} Clermont Auvergne, CNRS/IN2P3, LPC, Clermont-Ferrand, France\\
$^{126}$ Universit\'{e} de Lyon, CNRS/IN2P3, Institut de Physique des 2 Infinis de Lyon, Lyon, France\\
$^{127}$ Universit\'{e} de Strasbourg, CNRS, IPHC UMR 7178, F-67000 Strasbourg, France, Strasbourg, France\\
$^{128}$ Universit\'{e} Paris-Saclay Centre d'Etudes de Saclay (CEA), IRFU, D\'{e}partment de Physique Nucl\'{e}aire (DPhN), Saclay, France\\
$^{129}$ Universit\`{a} degli Studi di Foggia, Foggia, Italy\\
$^{130}$ Universit\`{a} del Piemonte Orientale, Vercelli, Italy\\
$^{131}$ Universit\`{a} di Brescia, Brescia, Italy\\
$^{132}$ Variable Energy Cyclotron Centre, Homi Bhabha National Institute, Kolkata, India\\
$^{133}$ Warsaw University of Technology, Warsaw, Poland\\
$^{134}$ Wayne State University, Detroit, Michigan, United States\\
$^{135}$ Westf\"{a}lische Wilhelms-Universit\"{a}t M\"{u}nster, Institut f\"{u}r Kernphysik, M\"{u}nster, Germany\\
$^{136}$ Wigner Research Centre for Physics, Budapest, Hungary\\
$^{137}$ Yale University, New Haven, Connecticut, United States\\
$^{138}$ Yonsei University, Seoul, Republic of Korea\\
$^{139}$  Zentrum  f\"{u}r Technologie und Transfer (ZTT), Worms, Germany\\
$^{140}$ Affiliated with an institute covered by a cooperation agreement with CERN\\
$^{141}$ Affiliated with an international laboratory covered by a cooperation agreement with CERN.\\

\end{flushleft}

\end{document}